\documentclass[prd,aps,preprintnumbers, showpacs, nofootinbib,superscriptaddress, notitlepage, 10pt]{revtex4-1}
\usepackage{epsfig}
\usepackage{bm}
\usepackage{amssymb}
\usepackage{amsmath}
\usepackage{color}
\usepackage{subcaption}
\usepackage[colorlinks,
            linkcolor=blue,
            anchorcolor=black,
            citecolor=blue
            ]{hyperref}
\newcommand{\beq}{\begin{eqnarray}}
\newcommand{\eeq}{\end{eqnarray}}
\newcommand{\nn}{\nonumber \\}
\newcommand{\ma}{\mathrm}

\begin{document}
\title{QCD evolution of the orbital angular momentum of quarks and gluons:\\ Genuine twist-three part}

\author{Yoshitaka Hatta}
\affiliation{Physics Department, Building 510A, Brookhaven National Laboratory, Upton, NY 11973, USA}

\author{Xiaojun Yao}
\affiliation{Department of Physics, Duke University, Durham, NC 27708, USA}

\begin{abstract}
We present the numerical solution of the one-loop QCD evolution equation for the genuine twist-three part of the orbital angular momentum (OAM) distributions  of quarks and gluons inside a longitudinally polarized nucleon. This is based on the observation that the evolution is identical to that of the Efremov-Teryaev-Qiu-Sterman function for transverse single spin asymmetry. Together with the known evolution of the Wandzura-Wilczek part,  the one-loop evolution of OAM distributions is now practically under control. We also study, for the first time, the scale dependence of the potential angular momentum defined as the difference between the Ji and Jaffe-Manohar definitions of OAM.

\end{abstract}
%\pacs{24.85.+p, 12.38.Bx, 12.39.St, 12.38.Cy}
\maketitle

\section{Introduction}

After more than a decade of experimental programs at RHIC, HERMES, COMPASS and JLab \cite{Adare:2008aa,Airapetian:2010ac,Adamczyk:2012qj,Alekseev:2010ub,Adamczyk:2014ozi,Prok:2014ltt}, and concurrent theoretical efforts based on global QCD analysis \cite{deFlorian:2014yva,Nocera:2014gqa,Sato:2016tuz}, we now know that the gluon helicity contribution $\Delta G$ to the Jaffe-Manohar sum rule of the nucleon spin \cite{Jaffe:1989jz}
\beq
\frac{1}{2}  =\frac{1}{2}\Delta \Sigma + \Delta G + L_{q} + L_g\,, \label{su}
\eeq
is nonzero and can be significant. Together with the well-known quark helicity contribution $\Delta \Sigma$, the first two terms of (\ref{su}) account for $\sim 70\%$ of the total spin, leaving some room for the orbital angular momentum (OAM) $L_{q,g}$ of quarks and gluons.  However, this is at best a tentative conclusion. The present estimate of $\Delta G$  is based on the RHIC 200 GeV $pp$ data which can only access a limited range of the Bjorken variable $x$. Uncertainties coming from the  small $x$-region $x\lesssim 0.05$ are estimated to be huge, and might completely alter the current estimates. In such circumstances, an obvious direction to proceed for the QCD spin community is to better constrain $\Delta G(x)$ in the small-$x$ region. Indeed,  new data from RHIC 510 GeV $pp$ runs have just begun to appear \cite{Adam:2019aml}. Moreover, the precise determination of $\Delta G$ has been designated as one of the primary goals of the future Electron-Ion Collider (EIC) \cite{Accardi:2012qut}.    

An equally obvious direction is to directly constrain the OAM contributions $L_{q,g}$. Unfortunately, however, this is currently not pursued experimentally due to the lack of clean observables from which one can systematically extract $L_{q,g}$ as the moment of the corresponding $x$-distributions $L_{q,g}=\int_0^1 dx L_{q,g}(x)$. In fact, it is only relatively recently that the rigorous definitions  of $L_{q,g}(x)$ have been derived  \cite{Hatta:2012cs}  (following the earlier developments \cite{Lorce:2011kd,Hatta:2011ku,Lorce:2011ni}),  although their gauge non-invariant versions have been known for quite some time \cite{Harindranath:1998ve,Hagler:1998kg}. Such a progress has finally led theorists to identify a few experimental processes that are sensitive to $L_{q,g}(x)$  \cite{Courtoy:2013oaa,Ji:2016jgn,Hatta:2016aoc,Bhattacharya:2017bvs,Bhattacharya:2018lgm}. In these works, observables have been computed to leading order, whereas a realistic global analysis requires the knowledge of the next-to-leading order (NLO) corrections together with the `DGLAP'  equation for $L_{q,g}(x)$ to NLO, or at least to leading order.  

However, so far the evolution of $L_{q,g}(x)$ has not been fully understood even to leading order.  This is because $L_{q}(x)$ and $L_g(x)$ are not the usual twist-two parton distribution functions (PDFs). They consist of the Wandzura-Wilczek (WW) part  and the genuine twist-three part 
\beq
L_{q,g}(x) = L_{q,g}^{WW}(x)+L_{3q,3g}(x)\,. \label{formula}
\eeq
The WW part is related to the twist-two unpolarized and polarized PDFs. As such, its evolution is known to the same order as that of the corresponding twist-two distributions (namely, to three-loops). The LO evolution equation derived in the early literature   
 \cite{Harindranath:1998ve,Hagler:1998kg} can be fully understood in this way \cite{Hoodbhoy:1998yb,Boussarie:2019icw}. More recently, the small-$x$ behavior of the WW part has received a lot of attention, mainly motivated by the aforementioned uncertainties of $\Delta G$ at small-$x$  \cite{Hatta:2018itc,More:2017zqp,Kovchegov:2019rrz,Boussarie:2019icw}. An analysis based on the double logarithmic resummation \cite{Boussarie:2019icw} has shown that there is a significant cancellation between $\Delta G(x)$ and $L_g(x)$ in the small-$x$ region (see, however, \cite{Kovchegov:2019rrz}).

On the other hand, the genuine twist-three part $L_{3q,3g}(x)$ is given by the matrix elements of three-parton ($q\bar{q}g,ggg$), twist-three operators. The QCD evolution of such twist-three correlators is complicated even to one-loop, and to our knowledge,  it has never been discussed  in the context of OAM. Fortunately, however, it has been discussed in a different context---QCD evolution for the Efremov-Teryaev-Qiu-Sterman (ETQS) function \cite{Efremov:1984ip,Qiu:1991wg} for transverse single spin asymmetry \cite{Kang:2008ey,Vogelsang:2009pj,Braun:2009mi}. Moreover, a C++ program to numerically solve this evolution has been developed by Pirnay \cite{Pirnay:2013fra}.  These results can be adapted for the present purpose with appropriate modifications. Thus, the goal of this paper is to articulate  the evolution equation for $L_{3q,3g}(x)$ and demonstrate the feasibility of numerically solving this equation.

The paper is organized as follows: In Sec.~\ref{sec2} we will explain the quark and gluon OAM distributions and their evolutions. Then in Sec.~\ref{sec3} the evolution of the genuine twist-three part will be discussed. We will show numerical results on the evolution of the genuine twist-three part in Sec.~\ref{sec4}. We will discuss the cusp anomalous dimensions for the large-$n$ moments and the scale dependence of the potential angular momentum. Finally, conclusions will be drawn in Sec.~\ref{sec5}.

% the  evolution equation is identical to the one derived in \cite{Braun:2009mi} (see, also, \cite{Kang:2008ey,Vogelsang:2009pj,Schafer:2012ra,Ma:2012xn,Yoshida:2016tfh}) in the context of transverse single spin asymmetry (SSA)  because the relevant operators involved are identical. Moreover, this  equation has been implemeted in a C++ code by Pirnay \cite{Pirnay:2013fra} and is publicly available from the arXiv webpage. In this paper, we shall adapt  this code to our problem and numerically solve the evolution equation for the twist-three part $L_{q,g}^{twist-3}(x)$. Together with the earlier works for the WW part, this gives the complete one-loop evolution of the OAM distribution.  

\section{OAM distributions and their evolution equation}
\label{sec2}
In this section, we first  review the formula (\ref{formula}) derived in \cite{Hatta:2012cs} and then discuss its evolution equation. We employ the following normalization for the singlet quark (and antiquark) OAM distribution $L_q(x)=\sum_f (L_f(x)+L_{\bar{f}}(x))$ and the gluon OAM distribution $L_g(x)$ 
\beq
L_q = \int_0^1dx L_q(x) , \qquad L_g=\int_0^1dx L_g(x)\,.
\eeq
Our normalization here is  different from the one in \cite{Hatta:2012cs},  but it is the same as in  \cite{Hatta:2018itc,Boussarie:2019icw}. In the present normalization and notation,  the result of \cite{Hatta:2012cs} reads 
\beq
L_q(x) &=& x\int^1_x \frac{dx'}{x'} (\Sigma(x')+E_q(x')) -x\int^1_x \frac{dx'}{x'^2}\Delta \Sigma(x') \nn 
&&-x \int^1_x dx_1 \int_{-1}^1 dx_2 \left(\Phi_F(x_1,x_2)-\Phi_F(-x_1,-x_2)\right) {\cal P}\frac{3x_1-x_2}{x_1^2(x_1-x_2)^2}  \nn 
&&-x \int^1_x dx_1 \int_{-1}^1 dx_2 \left(\widetilde{\Phi}_F(x_1,x_2)-\widetilde{\Phi}_F(-x_1,-x_2)\right) {\cal P}\frac{1}{x_1^2(x_1-x_2)}\,,  \label{ev1}
\eeq
\beq
L_g(x)&=& x\int^1_x \frac{dx'}{x'} (G(x')+E_g(x')) -2x\int^1_x \frac{dx'}{x'^2}\Delta G(x') \nn 
&&+2x \int^1_x \frac{dx'}{x'^3}\int dX (\Phi_F(x_1,x_2) -\Phi_F(-x_1,-x_2)) \nn 
&&+ 4x\int^1_xdx_1\int^1_{-1} dx_2 \widetilde{M}_F(x_1,x_2){\cal P} \frac{1}{x_1^3(x_1-x_2)} \nn && 
+ 4x\int^1_xdx_1\int^1_{-1} dx_2 M_F(x_1,x_2){\cal P} \frac{2x_1-x_2}{x_1^3(x_1-x_2)^2}\,, \label{ev2}
\eeq
where $X=\frac{x_1+x_2}{2}$ and $x'$ in the second line of (\ref{ev2}) means $x_1-x_2$. The first line on the right hand side is the WW part $L_{q,g}^{WW}(x)$ and the rest is the genuine twist-three part $L_{3q,3g}(x)$. $\Sigma(x)=\sum_f (f(x)+\bar{f}(x))$ and $G(x)$ are the usual, unpolarized quark and gluon distributions. (Ref.~\cite{Hatta:2012cs} used the notation $H_g=xG$.)  $E_{q}(x)=\sum_f (E_f(x)+E_{\bar{f}}(x))$ and $E_g(x)$ are  the helicity-flip quark and gluon generalized parton distributions (GPDs) commonly called `GPD $E$'.  

\begin{figure}
\centering
\includegraphics[width=14cm]{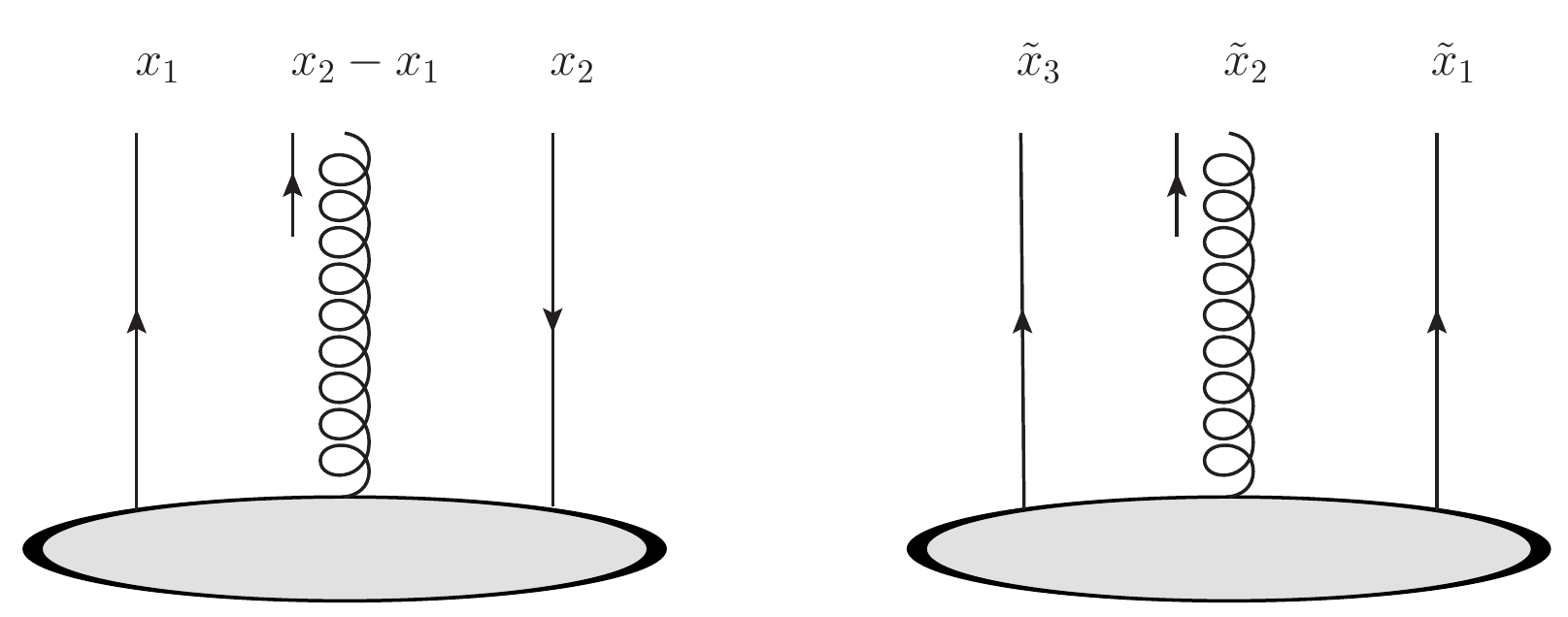}
\caption{Notations for the momentum fractions in Ref.~\cite{Hatta:2012cs} (left) and in Ref.~\cite{Braun:2009mi} (right).
}
\label{dif}
\end{figure}

The genuine twist-three distributions $\Phi_F$ and $\widetilde{\Phi}_F$ are defined through the matrix elements of quark-gluon correlation functions along the light-cone \cite{Hatta:2012cs,Ji:2012ba}
\beq
\int \frac{d\lambda d\zeta}{(2\pi)^2} e^{i\frac{\lambda}{2}(x_1+x_2) +i\zeta(x_2-x_1)} \langle P'S'|\bar{\psi}(-\lambda n/2)\gamma^+gF^{+i}(\zeta n)\psi(\lambda n/2)|PS\rangle = P^+\epsilon^{+i\rho\sigma}S_\rho \Delta_\sigma \Phi_F(x_1,x_2)+\cdots\,, \label{fi}
\eeq
%\beq
%\int \frac{d\lambda d\zeta}{(2\pi)^2} e^{i\frac{\lambda}{2}(x_1+x_2) +i\zeta(x_2-x_1)} \langle P'S'|\bar{\psi}(-\lambda n/2)\gamma^+ (-i\gamma_5) g\widetilde{F}^{+i}(\zeta n)\psi(\lambda n/2)|PS\rangle = P^+\epsilon^{+i\rho\sigma}S_\rho \Delta_\sigma \widetilde{\Phi}_F(x_1,x_2)+\cdots\,,  \label{se}
%\eeq
\beq
\int \frac{d\lambda d\zeta}{(2\pi)^2} e^{i\frac{\lambda}{2}(x_1+x_2) +i\zeta(x_2-x_1)} \langle P'S'|\bar{\psi}(-\lambda n/2)\gamma^+ (-i\gamma_5) gF^{+i}(\zeta n)\psi(\lambda n/2)|PS\rangle = P^+S^+ \Delta^i \widetilde{\Phi}_F(x_1,x_2)+\cdots\,,  \label{se}
\eeq
where $S^\mu=\delta^\mu_+S^+$ is the longitudinally polarized nucleon spin vector and $n^\mu=\delta^\mu_-$ is a fixed light-like vector.  $i,j=1,2$ denote transverse indices. The Wilson lines are omitted for simplicity. The summation over quark flavors is implied. The momentum transfer  $\Delta^\mu=P'^\mu-P^\mu = \delta^\mu_i \Delta^i$ is assumed to be small, has only transverse components, and we have kept only the linear terms in $\Delta$.  
The variables $1>x_{1,2}>-1$ can be interpreted as follows: The outgoing quark ($\psi$) and gluon ($F^{+i}$) carry momentum fractions $x_1$ and $x_2-x_1$ of the parent nucleon, respectively, and the returning quark ($\bar{\psi}$) has momentum fraction $x_2$, see Fig.~\ref{dif}. 
Similarly, for the three-gluon correlators $M_F$ and $\widetilde{M}_F$, one has 
\beq
\int \frac{d\lambda d\zeta}{(2\pi)^2} e^{i\frac{\lambda}{2}(x_1+x_2) +i\zeta(x_2-x_1)} \langle P'S'|F^{+\alpha}(-\lambda n/2)gF^{+i}(\zeta n)F^{+}_{\ \alpha}(\lambda n/2)|PS\rangle = (P^+)^2\epsilon^{+i\rho\sigma}S_\rho \Delta_\sigma M_F(x_1,x_2)+\cdots\,,
\eeq
%\beq \nonumber
%\int \frac{d\lambda d\zeta}{(2\pi)^2} e^{i\frac{\lambda}{2}(x_1+x_2) +i\zeta(x_2-x_1)} \langle P'S'|\epsilon_{ik}F^{+i}(-\lambda n/2)gF^{+j}(\zeta n)F^{+k}(\lambda n/2)|PS\rangle = (P^+)^2\epsilon_i^{\ j}\epsilon^{+i\rho\sigma}S_\rho \Delta_\sigma \widetilde{M}_F(x_1,x_2)+\cdots\,, \\
%\label{cc}
%\eeq
\beq 
\int \frac{d\lambda d\zeta}{(2\pi)^2} e^{i\frac{\lambda}{2}(x_1+x_2) +i\zeta(x_2-x_1)} \langle P'S'|\epsilon_{ik}F^{+i}(-\lambda n/2)gF^{+j}(\zeta n)F^{+k}(\lambda n/2)|PS\rangle = -(P^+)^2S^+\Delta^j \widetilde{M}_F(x_1,x_2)+\cdots,
\label{cc}
\eeq
where  $FFF$ means  $F_a (T^b)_{ac}F_b F_c  =if_{abc}F_a F_b F_c$ in color space. 
% This is the same as the notation by Kang and Qiu, e.g., $\Phi_F(x_1,x_2) \sim \widetilde{{\cal T}}_{q,F}(x_1,x_2)$. 
These distributions have the following symmetry properties
\beq
\Phi_F(x_1,x_2)=-\Phi_F(x_2,x_1)\,, \qquad \widetilde{\Phi}_F(x_1,x_2)=\widetilde{\Phi}_F(x_2,x_1)\,, \nn
M_F(x_1,x_2)=-M_F(x_2,x_1)\,, \qquad \widetilde{M}_F(x_1,x_2)=\widetilde{M}_F(x_2,x_1) \,. \label{sym}
\eeq

Note that the total (quark plus gluon) genuine twist-three OAM distribution integrates to zero
 \beq
 \int_0^1 dx (L_{3q}(x) + L_{3g}(x))=0. \label{sum}
 \eeq
 We can write this as 
 \beq
\int_0^1 dx L_{3g}(x) \equiv L_\ma{pot} = - \int_0^1 dx L_{3q}(x) . \label{pot}
\eeq
The quantity $L_\ma{pot}$,  sometimes called the potential angular momentum  \cite{Wakamatsu:2010qj},  represents the difference between  the kinetic (Ji) OAM \cite{Ji:1996ek} and the canonical (Jaffe-Manohar) OAM \cite{Jaffe:1989jz}
\beq
L_\ma{Ji}^q - L_\ma{JM}^q = L_\ma{pot}.
\eeq
Roughly speaking, it arises from the difference between  the covariant  derivative (kinetic momentum) $\vec{x}\times \vec{D}$ and the partial derivative (canonical momentum) $\vec{x}\times \vec{\partial}$ in the definition of OAM. The value of $L_\ma{pot}$ for an electron in QED has been the subject of debate in the literature \cite{Burkardt:2008ua,Liu:2014fxa,Ji:2015sio}. More recently,  $L_\ma{pot}$ has been calculated in lattice QCD and found to be nonzero \cite{Engelhardt:2017miy,Engelhardt:2019lyy}.  

Let us now consider the evolution of $L_{q,g}(x)$. Taking the $n$-th moments $L_{q,g}^n=\int_0^1 dx x^{n-1}L_{q,g}(x)$, etc., we find
\beq
L_{q}^n&=& \frac{1}{n+1}(\Sigma^{n+1} +E_q^{n+1}) -\frac{1}{n+1}\Delta\Sigma^n+L^n_{3q}, \nn 
L_g^n&=& \frac{1}{n+1}(G^{n+1}+E_g^{n+1})-\frac{2}{n+1}\Delta G^n +L^n_{3g} \,,\label{coin}
\eeq
where $L_{3q}$ and $L_{3g}$ are  the contributions from the genuine twist three part. 
 The evolution of $\Sigma^n,G^n,\Delta\Sigma^n$ and $\Delta G^n$  in the renormalization scale $\mu^2$ is governed by the standard DGLAP anomalous dimensions $\gamma^n$ and $\Delta \gamma^n$. Moreover, the anomalous dimensions of  $E^n_q,E^n_g$ are the same as those for  $\Sigma^n,G^n$. Therefore, if one restricts oneself to the WW part, one can immediately write down the evolution equation  
\beq
\frac{\partial}{\partial t} \begin{pmatrix} L^{n}_q \\ L_g^{n} \end{pmatrix}^{WW} &=& \frac{1}{n+1} \begin{pmatrix} \gamma^{n+1}_{qq} & \gamma^{n+1}_{qg} \\ \gamma^{n+1}_{gq} & \gamma^{n+1}_{gg} \end{pmatrix} \begin{pmatrix} \Sigma^{n+1}+E_q^{n+1} \\ G^{n+1}+E_g^{n+1} \end{pmatrix} -\frac{1}{n+1} \begin{pmatrix} \Delta \gamma^{n}_{qq} & \Delta \gamma^n_{qg} \\ 2\Delta \gamma^n_{gq} & 2\Delta \gamma^n_{gg} \end{pmatrix}  \begin{pmatrix} \Delta \Sigma^n \\ \Delta G^n \end{pmatrix}   \nn 
&=&  \begin{pmatrix} \gamma^{n+1}_{qq} & \gamma^{n+1}_{qg} \\ \gamma^{n+1}_{gq} & \gamma^{n+1}_{gg} \end{pmatrix}  \begin{pmatrix} L^n_q \\ L_g^n \end{pmatrix}^{WW}      +\frac{1}{n+1} \begin{pmatrix} \gamma_{qq}^{n+1}- \Delta \gamma^{n}_{qq} & 2\gamma_{qg}^{n+1}- \Delta \gamma^n_{qg} \\ \gamma_{gq}^{n+1}-2\Delta \gamma^n_{gq} & 2\gamma_{gg}^{n+1}- 2\Delta \gamma^n_{gg} \end{pmatrix}  \begin{pmatrix} \Delta \Sigma^n \\ \Delta G^n \end{pmatrix} \,,
 \label{wa}
 \eeq
 where 
 \beq
 t= - \frac{2}{b_0} \ln \frac{\alpha_s(\mu^2)}{\alpha_s(\mu_0^2)}, \label{t}
 \eeq
 with  $\alpha_s(\mu^2) = \frac{4\pi}{b_0\ln (\mu^2/\Lambda_{QCD}^2)}$ and $b_0=\frac{11N_c}{3}-\frac{2n_f}{3}$. 
 In the WW approximation, (\ref{wa}) holds to all orders in perturbation theory \cite{Boussarie:2019icw}. To one-loop order, it agrees with the result obtained in \cite{Harindranath:1998ve,Hagler:1998kg}, see also, \cite{Hoodbhoy:1998yb}.

Inclusion of the genuine twist-three parts makes things considerably more complicated. One might naively expect that the genuine twist-three part would evolve with its own anomalous dimensions
\beq
\frac{\partial}{\partial t} \begin{pmatrix} L^n_{3q} \\ L_{3g}^n \end{pmatrix} \stackrel{?}{=}
\begin{pmatrix} A_n & B_n \\ C_n & D_n \end{pmatrix} \begin{pmatrix} L_{3q}^n \\ L_{3g}^n \end{pmatrix} \,, \label{naive}
\eeq
and therefore the evolution of the full OAMs would be given by
\beq
\frac{\partial}{\partial t} \begin{pmatrix} L^n_q \\ L_g^n \end{pmatrix} &\stackrel{?}{=}&  \begin{pmatrix} \gamma^{n+1}_{qq} & \gamma^{n+1}_{qg} \\ \gamma^{n+1}_{gq} & \gamma^{n+1}_{gg} \end{pmatrix}  \begin{pmatrix} L^n_q \\ L_g^n \end{pmatrix}      +\frac{1}{n+1} \begin{pmatrix} \gamma_{qq}^{n+1}- \Delta \gamma^{n}_{qq} & 2\gamma_{qg}^{n+1}- \Delta \gamma^n_{qg} \\ \gamma_{gq}^{n+1}-2\Delta \gamma^n_{gq} & 2\gamma_{gg}^{n+1}- 2\Delta \gamma^n_{gg} \end{pmatrix}  \begin{pmatrix} \Delta \Sigma^n \\ \Delta G^n \end{pmatrix}  \nn &&+
\begin{pmatrix} A_n-\gamma_{qq}^{n+1} & B_n -\gamma_{qg}^{n+1} \\ C_n-\gamma_{gq}^{n+1} & D_n -\gamma^{n+1}_{gg} \end{pmatrix} \begin{pmatrix} L_{3q}^n \\ L_{3g}^n \end{pmatrix} \,. \label{naive2}
 \eeq
However, we shall demonstrate later that this is not the case.  In general, different moments (different $n$'s) mix under evolution so that the right hand side of  (\ref{naive}) should involve a summation over all moments. This suggests that it is more convenient to study the evolution directly in the $x$-space.

\section{Evolution of the genuine-twist three part}
\label{sec3}
At first sight, the $\mu^2$-evolution of $\Phi_F(x_1,x_2)$ etc., hence that of $L_{3q,3g}(x)$ seems a challenging open question. However, it is actually known in the  context of transverse single spin asymmetry (SSA). There, exactly the same set of operators as in  (\ref{fi})-(\ref{cc}) appear. The difference is that in the case of SSA, one takes the forward matrix element (i.e., $\Delta=0$) in the transversely polarized nucleon state $S^\mu=\delta^\mu_i S_i$. The resulting distribution, the Efremov-Teryaev-Qiu-Sterman (ETQS) function \cite{Efremov:1984ip,Qiu:1991wg} has been extensively discussed in the literature. Because the evolution is intrinsic to the operators involved, not to external states, and here we only consider the $\Delta\to 0$ limit of the nonforward matrix element,\footnote{ In principle, $\Phi_F$ etc. depend on $\Delta^2$ but this dependence has been neglected in (\ref{fi}) since it is of higher order.}  the same evolution equation should apply. The complete derivation of the evolution equation has been given in   \cite{Braun:2009mi} following earlier attempts   \cite{Kang:2008ey,Vogelsang:2009pj} (see also \cite{Schafer:2012ra,Ma:2012xn,Yoshida:2016tfh}).  The result is too lengthy to be reproduced here, but fortunately, a C++ code is publicly available  \cite{Pirnay:2013fra}. We shall heavily rely on this code in the following.

 For this purpose, first we need to clarify the difference in notations between ours and in  \cite{Braun:2009mi,Pirnay:2013fra}. There the authors introduced the following  distributions with three arguments
\beq
T_{\bar{q}Fq}(\tilde{x}_1,\tilde{x}_2,\tilde{x}_3), \qquad \Delta T_{\bar{q}Fq}(\tilde{x}_1,\tilde{x}_2,\tilde{x}_3), \quad T_{3F}^+(\tilde{x}_1,\tilde{x}_2,\tilde{x}_3), \quad  \Delta T_{3F}^+(\tilde{x}_1,\tilde{x}_2,\tilde{x}_3) \,. \label{four}
\eeq
 To avoid confusion, we have added a tilde on momentum fractions. Their meaning is self-explanatory from  Fig.~\ref{dif} (beware of the direction of arrows) with the correspondence
  \beq
   x_1=\tilde{x}_3, \quad x_2-x_1=\tilde{x}_2 , \quad  x_2=-\tilde{x}_1, \qquad  \tilde{x}_1+\tilde{x}_2+\tilde{x}_3=0 \,.
 \eeq
 The four distributions in (\ref{four}) are direct analogs of (\ref{fi})-(\ref{cc}). The plus sign on $T_{3F}$ and $\Delta T_{3F}$ means the contraction of color indices with the $f$-symbol, cf., the comment below (\ref{cc}). We have carefully checked the relative normalization and found the correspondence\footnote{ This comparison is complicated by the fact that Refs.~\cite{Hatta:2012cs,Braun:2009mi} use different conventions for $\gamma_5$ and the epsilon tensor.  Ref.~\cite{Hatta:2012cs} used $\epsilon^{0123}=+1$ whereas  Ref.~\cite{Braun:2009mi} used $\epsilon_{0123}=-\epsilon^{0123}=1$. The sign of $\gamma_5$ is also opposite in the two references. 
}  
 \beq
\Phi_F(x_1,x_2) \leftrightarrow 2T_{\bar{q}Fq}(-x_2,x_2-x_1,x_1), \label{E1}
\eeq
\beq
\widetilde{\Phi}_F(x_1,x_2) \leftrightarrow -2\Delta T_{\bar{q}Fq}(-x_2,x_2-x_1,x_1), \label{E2}
\eeq
\beq
M_F(x_1,x_2) \leftrightarrow T^+_{3F} (-x_2,x_2-x_1,x_1), \label{E3}
\eeq
 \beq
\widetilde{M}_F(x_1,x_2)  \leftrightarrow -\Delta T^+_{3F}( \tilde{x}_1,\tilde{x}_2,\tilde{x}_3) \,.\label{E4}
\eeq
As observed in \cite{Braun:2009mi}, $\Delta T^+_{3F}$ is not an independent function (see Eq.~(32) there). Accordingly, $\widetilde{M}_F$ can be eliminated via the corresponding formula
 \beq
 \widetilde{M}_F(x_1,x_2) =-M_F(x_2-x_1,x_2) + M_F(x_1,x_1-x_2). \label{re}
 \eeq
This  was not noticed in  \cite{Hatta:2012cs}, but it can be indeed derived from the formulas in this reference. 
 %This is a direct analog of  (32) (upper sign). To see this, note that $(\tilde{x}_1,\tilde{x}_2,\tilde{x}_3)\to (\tilde{x}_1,\tilde{x}_3,\tilde{x}_2)$ is equivalent to replacing $x_1\to x_2-x_1$, and $(\tilde{x}_1,\tilde{x}_2,\tilde{x}_3)\to (\tilde{x}_2,\tilde{x}_1,\tilde{x}_3)$ is equivalent to $x_2\to x_1-x_2$. This is why Pirnay's code does not provide an output for $\Delta T_{3F}$; it is missing in Table 1.  

Another complication is that in Ref.~\cite{Braun:2009mi}, the evolution equation has been  presented in terms of 
\beq
\mathfrak{S}^\pm (\tilde{x}_1,\tilde{x}_2,\tilde{x}_3), \qquad \mathfrak{F}^\pm(\tilde{x}_1,\tilde{x}_2,\tilde{x}_3)
\eeq
which are particular linear combinations of the distributions in (\ref{four}).  
The signs $\pm$ refer to $C$-parity, and the point is that the evolution equation does not mix the $C$-even and $C$-odd functions.  From (\ref{E1})-(\ref{E4}), we find the correspondence
\beq
2\mathfrak{S}^+ (\tilde{x}_1,\tilde{x}_2,\tilde{x}_3 )&\leftrightarrow& \Phi_F(x_1,x_2)-\Phi_F(-x_1,-x_2) + \widetilde{\Phi}_F(x_1,x_2)-\widetilde{\Phi}_F(-x_1,-x_2) 
\nn && \equiv  2S^+(x_1,x_2) \,. \label{im1}
%-2S^+(-x_1,-x_2)
\eeq
\beq
\mathfrak{F}^+(\tilde{x}_1,\tilde{x}_2,\tilde{x}_3) &\leftrightarrow&  M_F(x_1,x_2) + \widetilde{M}_F(x_1,x_2) \nn && \equiv F^+(x_1,x_2) \,. \label{im2}
% =  F^+(-x_1,-x_2)
\eeq
The distributions $\mathfrak{S}^+$ and $\mathfrak{F}^+$ have the following symmetry \cite{Braun:2009mi}
\beq
\mathfrak{S}^+(-\tilde{x})=\mathfrak{S}^+(\tilde{x}), \qquad \mathfrak{F}^+(-\tilde{x})=-\mathfrak{F}^+(\tilde{x}). \label{s1}
\eeq
In our case, we find from (\ref{sym})\footnote{ In the three-gluon sector, there is another relation 
\beq
F^+(x_1,x_2)=-F^+(x_2-x_1, x_2) \,. \label{s2}
\eeq
This has the same sign as the corresponding relation $\mathfrak{F}^+(\tilde{x}_1,\tilde{x}_2,\tilde{x}_3)=-\mathfrak{F}^+(\tilde{x}_1,\tilde{x}_3,\tilde{x}_2)$ in \cite{Braun:2009mi} (see the unnumbered equation below (30)). There is no contradiction here. } 
\beq
S^+(-x_1,-x_2)=-S^+(x_1,x_2), \qquad F^+(-x_1,-x_2)=F(x_1,x_2) .
\eeq
Notice the differences in sign, which is simply due to the fact that we are dealing with different matrix elements (longitudinally polarized nucleon, finite momentum transfer). This however causes no problem in practice because the evolution equation automatically preserves the symmetry property of the initial conditions.

Note also that in (\ref{im1}) and (\ref{im2}), we only showed the $C$-even distributions. This is because the OAM distributions are $C$-even ($PT$-even) \cite{Hatta:2011ku}. 
 Indeed, the twist-three part of (\ref{ev1}) and (\ref{ev2}) can be solely written in terms of $C$-even functions $S^+$ and $F^+$. It is not difficult to show that
\beq
%L_\Sigma(x)&=& x\int^1_x \frac{dx'}{x'} (\Sigma(x')+E_\Sigma(x')) -x\int^1_x \frac{dx'}{x'^2}\Delta \Sigma(x') + L_{3\Sigma}(x) \\
L_{3q}(x)  &=& -2x \int^1_x dx_1 \int_{-1}^1 dx_2 S^+(x_1,x_2) \frac{1}{x_1^2(x_1-x_2)}  \nn
&& -2x \int^1_x 
dx_1 \int_{-1}^1 dx_2 (S^+(x_1,x_2)-S^+(x_2,x_1)) {\cal P}\frac{1}{x_1(x_1-x_2)^2} \,. \label{final1}
\eeq
In the gluon case, let us write\footnote{Note that  we can replace $\Phi_F(x_1,x_2)-\Phi_F(-x_1,-x_2)$ with $2S^+(x_1,x_2)$ because the $\widetilde{\Phi}_F$ part of $S^+$ drops out $\int dX (\widetilde{\Phi}_F(X,x)-\widetilde{\Phi}_F(-X,-x))=0$.}   
\beq
%L_g(x) &=& x\int^1_x \frac{dx'}{x'} (G(x')+E_g(x')) -2x\int^1_x \frac{dx'}{x'^2}\Delta G(x') + L_{3g}(x) \\
L_{3g}(x) &=& L_{3g,S}(x)  + L_{3g,F}(x), \label{d}
\eeq
where 
\beq
L_{3g,S}(x) &=& 4x \int^1_x \frac{dx'}{x'^3}\int dX  S^+(x_1,x_2) \\
L_{3g,F}(x) &=& 2x\int^1_xdx_1\int^1_{-1} dx_2 (F^+(x_1,x_2) -F^+(x_2,x_1)){\cal P} \frac{1}{x_1^2(x_1-x_2)^2} \nn && 
+ 4x\int^1_xdx_1\int^1_{-1} dx_2 F^+ (x_1,x_2){\cal P} \frac{1}{x_1^3(x_1-x_2)} \,. \label{in}
\eeq
In terms of moments, we have

\beq
L_{3q}^n % &=& \int_0^1 dx x^{n-1} I_\Sigma(x) \nn
= -\frac{2}{n+1} \int_0^1 dx_1  \int_{-1}^1 dx_2  \Biggl[S^+(x_1,x_2) \frac{x_1^{n-1}}{x_1-x_2} 
 + (S^+(x_1,x_2)-S^+(x_2,x_1)) {\cal P}\frac{x_1^{n}}{(x_1-x_2)^2}   \Biggr]\,.
\eeq
\beq
 L_{3g}^n %&=& \int_0^1 dx x^{n-1} I_g(x) \nn
 &=& \frac{4}{n+1} \int_0^1 dx \int_{-1}^1 dX x^{n-2}S^+\left(X+\frac{x}{2},X-\frac{x}{2}\right) \nn 
 && -\frac{2}{n+1} \int_0^1dx_1 \int_{-1}^1 dx_2 \Biggl[  (F^+(x_1,x_2) -F^+(x_2,x_1)){\cal P} \frac{x_1^{n-1}}{(x_1-x_2)^2}  
+ 2 F^+ (x_1,x_2){\cal P} \frac{x_1^{n-2}}{(x_1-x_2)}  \Biggr] \,.  \label{final4}
\eeq
Eqs.~(\ref{final1})-(\ref{final4}) are the starting point of our numerical analysis which we now turn to.

\section{Numerical results}
\label{sec4}
In this section, we present the one-loop scale evolution of the genuine twist-three part of the OAM distributions $L_{3q,3g}(x,\mu)$. Our numerical results are based on a C++ code developed by Pirnay  \cite{Pirnay:2013fra}.\footnote{Here we clarify a few things and list a few typos in the code:
 \begin{enumerate}
 \item The code in Ref.~\cite{Pirnay:2013fra} is based on the evolution kernels derived in Ref.~\cite{Braun:2009mi}. Ref.~\cite{Braun:2009mi} calculates the kernels with the evolution equation defined as $\frac{\partial}{\partial \ln \mu} \cdots$. The code in Ref.~\cite{Pirnay:2013fra} also implements the evolution equation numerically as $\frac{\partial}{\partial \ln \mu} \cdots$. But the manuscript of Ref.~\cite{Pirnay:2013fra} writes the evolution equation as $\frac{\partial}{\partial \ln \mu^2} \cdots$. The expression (52) in Ref.~\cite{Braun:2009mi} has a typo: the factor $\frac{\alpha_s}{4\pi}$ in the third term should be $\frac{\alpha_s}{2\pi}$.
 \item In the line 347 of the code file `t3evol.cpp', the `F+\_initial.txt' should be `F-\_initial.txt'.
 \item In the lines 65, 90, 117 and 142 of the code file `fffkernels.h', there should be no `Nc' multiplying `beta0(nf)'.
 \end{enumerate}
 We would like to thank Vladimir Braun and Alexander Manashov for confirming these.} Although the code is intended for the ETQS function, it can be straightforwardly adapted to the present problem with almost no change. The only thing one should keep in mind is that the symmetry properties of the various distributions are different, see (\ref{s1}) and (\ref{s2}). However, this does not cause extra complications since the evolution equation preserves this symmetry, and one just needs to properly implement the relevant symmetry in the initial conditions.

We set the number of lattice sites in the interval $-1\leq x \leq 1$ to be $N_x=2001$ (this has to be an odd number in the code), so that the lattice size is $\Delta x = 0.001$. As for the initial conditions, currently nothing is known about the functional forms of the genuine twist-three distributions $\Phi(x_1,x_2)$, etc., except that they must vanish at kinematic boundaries (e.g., $x_1=\pm 1$) and obey the symmetry properties (\ref{sym}). We thus introduce three different models at the initial scale $\mu_0^2$
\beq
\label{eqn:init_phiF}
\Phi_F(x_1, x_2) &=& \begin{cases} 
			0, & |x_1| \geq 1 \parallel |x_2| \geq 1 \parallel |x_2 - x_1| \geq 1 \\
			2(x_1-x_2)f(x_1, x_2) , & \ma{otherwise} \end{cases}\\ 
\label{eqn:init_tildephiF}
\widetilde{\Phi}_F(x_1, x_2) &=& \begin{cases} 
			0, & |x_1| \geq 1 \parallel |x_2| \geq 1 \parallel |x_2 - x_1| \geq 1 \\ 
			2f(x_1, x_2) , & \ma{otherwise} \end{cases}\\
\label{eqn:init_MF}
M_F(x_1, x_2) &=& \begin{cases} 
			0, & |x_1| \geq 1 \parallel |x_2| \geq 1 \parallel |x_2 - x_1| \geq 1 \\ 
			(x_1^2-x_2^2)f(x_1, x_2) , & \ma{otherwise} \end{cases} 
\eeq
%We assume the initial conditions for both the flavor singlet and nonsinglet are given by (\ref{eqn:init_phiF}) and (\ref{eqn:init_tildephiF}). We will focus on the evolution of the flavor singlet and the gluon parts in most of the following studies.
with
\begin{enumerate}
\item $f(x_1, x_2) = \frac{1}{10}(1-x_1^2)(1-x_2^2)(1-(x_1-x_2)^2)$\,;
\item $f(x_1, x_2) = x_1x_2(1-x_1^2)(1-x_2^2)(1-(x_1-x_2)^2)$\,;
\item $f(x_1, x_2) = x^2_1x^2_2(1-x_1^2)(1-x_2^2)(1-(x_1-x_2)^2)$\,.
\end{enumerate}
The overall sign has been fixed such that the potential angular momentum (\ref{pot}) calculated from these initial conditions is positive $L_\ma{pot}>0$, as suggested by an intuitive argument \cite{Burkardt:2012sd} as well as a recent lattice calculation   \cite{Engelhardt:2017miy}. 
We then use (\ref{im1}) and (\ref{im2}) to convert the above initial conditions  into those of $S^+$ and $F^+$. They are further converted into $\mathfrak{S}^+(\tilde{x}_1,\tilde{x}_2,\tilde{x}_3)=S^+(x_1,x_2)$ and $\mathfrak{F}^+(\tilde{x}_1,\tilde{x}_2,\tilde{x}_3)=F^+(x_1,x_2)$ since the code requires $\mathfrak{S}^+$ and $\mathfrak{F}^+$ as the initial conditions.  In the end of the numerical calculation, we convert the final results of $\mathfrak{S}^+$ and $\mathfrak{F}^+$ back into $S^+$ and $F^+$. Finally, we compute the twist-three OAM distributions $L_{3q}(x) $ and $L_{3g}(x)$, Eqs.~(\ref{final1})-(\ref{in}), numerically by using the Simpson's rule. The integrands of $L_{3q}(x) $ and $L_{3g}(x)$ may have singularities when $x_1\to0$ or $x'\to0$. Such singularities may be present already in the initial conditions, or can be dynamically generated by the evolution (see below). 
To improve the numerical accuracy when this happens, we first fit the integrand $f(x_1)$ (or $f(x')$) in the small-$x$ region (in practice, $x<0.005$) by the power function $ax_1^b$ (or $ax'^b$) with $b<0$. Then we split as
\beq
\int_x^1 dx_1 f(x_1) = \int_x^1 dx_1 (f(x_1)-ax_1^b) + \int_x^1 dx_1 ax_1^b.
\eeq
The integrand of the first term has no singularities so the integral can be evaluated accurately. The second integral can be done analytically. 
%integrate the original integrand subtracted by the fitted power function, which has no singularities when $x_1\to0$ ($x'\to0$). The result of the original integration is given by the sum of the numerical integration value and the integration value of the power function, which can be done analytically. 

As a warm-up, let us first check the sum rules  \cite{Hatta:2012cs}:
\beq
\label{eqn:sum1}
\int_0^1 dx L_{3g,F}(x) &=& 0, \\
\label{eqn:sum2}
\int_0^1 dx ( L_{3q}(x)  + L_{3g,S}(x) ) &=& 0 \,.
\eeq
This is a more detailed version of (\ref{sum}) (see (\ref{d})). We numerically evolve each of the initial conditions from $\mu_0^2=1$ GeV$^2$ to $\mu^2=10$ GeV$^2$.
The numerical results of the left-handed-sides of (\ref{eqn:sum1}) and (\ref{eqn:sum2}) for the three different initial conditions are shown in Table~\ref{table:1}. As a reference, we also show the value of the potential angular momentum $L_\ma{pot}$ (\ref{pot}) in these models. We see that the sum rules are satisfied up to numerical errors\footnote{The default time step in $t$ (see (\ref{t})) in the evolution code is $\Delta t=0.01$. However, we find it necessary to make this 0.001 for the initial condition 1, because a singularity is developed at small-$x$ for this initial condition and we need a higher accuracy to stabilize the evolution and satisfy the sum rules. 
% When we evolve the initial condition 1 from $\mu_0^2=1$ GeV$^2$ to $\mu^2=10$ GeV$^2$, we change the time step to be $0.001$. Otherwise, the sum rule integrals are divergent. 
We also use $0.001$ as the time step for initial conditions 2 and 3.}.

\begin{table*}[ht]
\caption{\label{table:1}Numerical results of the sum rules.}
\begin{center}
\begin{tabular}{|c|c|c|c|}
\hline
  & Initial condition 1 & Initial condition 2 & Initial condition 3 \\
\hline
$\int_0^1 dx L_{3g,F}(x)$ at $\mu_0^2=1$ GeV$^2$ &  $0.0006$ & $0.0006$ &  $0.0002$  \\
%\hline
%$\mu^2=1.01$ GeV$^2$ &  $0.004$ & $0.0006$ &  $0.0002$  \\
\hline
$\int_0^1 dx L_{3g,F}(x)$ at $\mu^2=10$ GeV$^2$ &  -0.00007 & 0.0004 &  0.00016 \\
\hline
$\int_0^1 dx ( L_{3q}(x)  + L_{3g,S}(x) )$ at $\mu_0^2=1$ GeV$^2$ &  $4\times10^{-7}$ & -0.0006 &  -0.0002  \\
%\hline
%$\mu^2=1.01$ GeV$^2$ &  $0.004$ & $-4\times10^{-5}$ & $-1\times10^{-5}$  \\
\hline
$\int_0^1 dx ( L_{3q}(x)  + L_{3g,S}(x) )$ at $\mu^2=10$ GeV$^2$ &  0.00014 & -0.0007 &  -0.0002  \\
\hline
$L_\ma{pot} $ at $\mu_0^2=1$ GeV$^2$ &  0.234 & 0.185 &  0.069  \\
\hline
$L_\ma{pot}$ at $\mu^2=10$ GeV$^2$ &  0.162 & 0.077 &  0.032  \\
\hline

\end{tabular}
\end{center}
\end{table*}

Next, we study whether the moments of the twist-three OAM distributions $L^n_{3q}$ (flavor singlet) and $L^n_{3g}$ satisfy the homogeneous equation (\ref{naive}). This can be tested by assuming (\ref{naive}) and considering an infinitesimal evolution 
\beq
\label{eqn:evol}
\begin{pmatrix} L^n_{3q} \\ L_{3g}^n \end{pmatrix}_{\mu^2} \approx  \begin{pmatrix} L^n_{3q} \\ L_{3g}^n \end{pmatrix}_{\mu_0^2} + \frac{2}{b_0}\ln \frac{\alpha_s(\mu^2_0)}{\alpha_s(\mu^2)} \begin{pmatrix} A_n & B_n \\ C_n & D_n \end{pmatrix}\begin{pmatrix} L^n_{3q} \\ L_{3g}^n \end{pmatrix}_{\mu^2_0} \,.
\eeq
% with  $\alpha_s(\mu^2) = \frac{4\pi}{b_0\ln \mu^2/\Lambda_{QCD}^2}$  and $b_0=11N_c/3-2n_f/3$.
In practice, we use $\mu_0^2=1$ GeV$^2$ and $\mu^2=1.01$ GeV$^2$. This means that $n_f=3$, and the code uses the value $\Lambda_{QCD}=0.204$ GeV in this case. In order to extract $A_n,B_n,...$, it is not enough to consider one initial condition since (\ref{eqn:evol}) contains two equations but there are four unknowns.  
% In the twist-three evolution code, $\Lambda_{QCD}=0.204$ GeV if $n_f=3$ and $0.175$ GeV if $n_f=4$.
 %For the evolution from $\mu_0^2=1$ GeV$^2$ to $\mu^2=1.01$ GeV$^2$, we set $n_f=3$. 
 %We cannot solve for $A_n$ and $B_n$ ($C_n$ and $D_n$) simultaneously by just using one set of initial conditions because the number of unknown parameters is two while the number of equations is just one. 
 But we can combine any two sets of initial conditions and form a solvable system of linear equations. We will label the results of $A_n,B_n,...$ obtained from the $i$-th and $j$-th initial conditions as $A_n(ij),B_n(ij)$,... and compare these with the known anomalous dimensions of the twist-two operators
\beq
\begin{pmatrix} \gamma^{n+1}_{qq} & \gamma^{n+1}_{qg} \\ \gamma^{n+1}_{gq} & \gamma^{n+1}_{gg} \end{pmatrix} =  \begin{pmatrix} \frac{4}{3}\left[-\frac{1}{2}+ \frac{1}{(n+1)(n+2)} -2\sum_{k=2}^{n+1} \frac{1}{k}\right] & n'_f \frac{ n^2+3n +4}{(n+1)(n+2)(n+3)} \\ 
\frac{4}{3} \frac{n^2+3n+4}{(n+1)n(n+2)} & 6\left[-\frac{1}{12}+\frac{1}{n(n+1)}+ \frac{1}{(n+2)(n+3)} -\sum_{k=2}^{n+1}\frac{1}{k} \right] -\frac{n'_f}{3} \end{pmatrix}  \,.         \label{st}    
\eeq
We make this comparison in the (naive) hope that there may be a significant cancelation in the second line of (\ref{naive2}) so that the evolution equation for the total OAM distributions is formally the same as that in the WW approximation. Note that we set $n'_f=2$ in (\ref{st}) because this is the value used in the one-loop evolution kernel in the code. It is different from $n_f$ used in the running coupling constant. The latter depends on the scale $\mu^2$. %This number of flavors is the one in the evolution and fixed to be $n'_f=2$ in the code. 

\begin{figure}
    \centering
    \begin{subfigure}[t]{0.48\textwidth}
        \centering
        \includegraphics[height=2.2in]{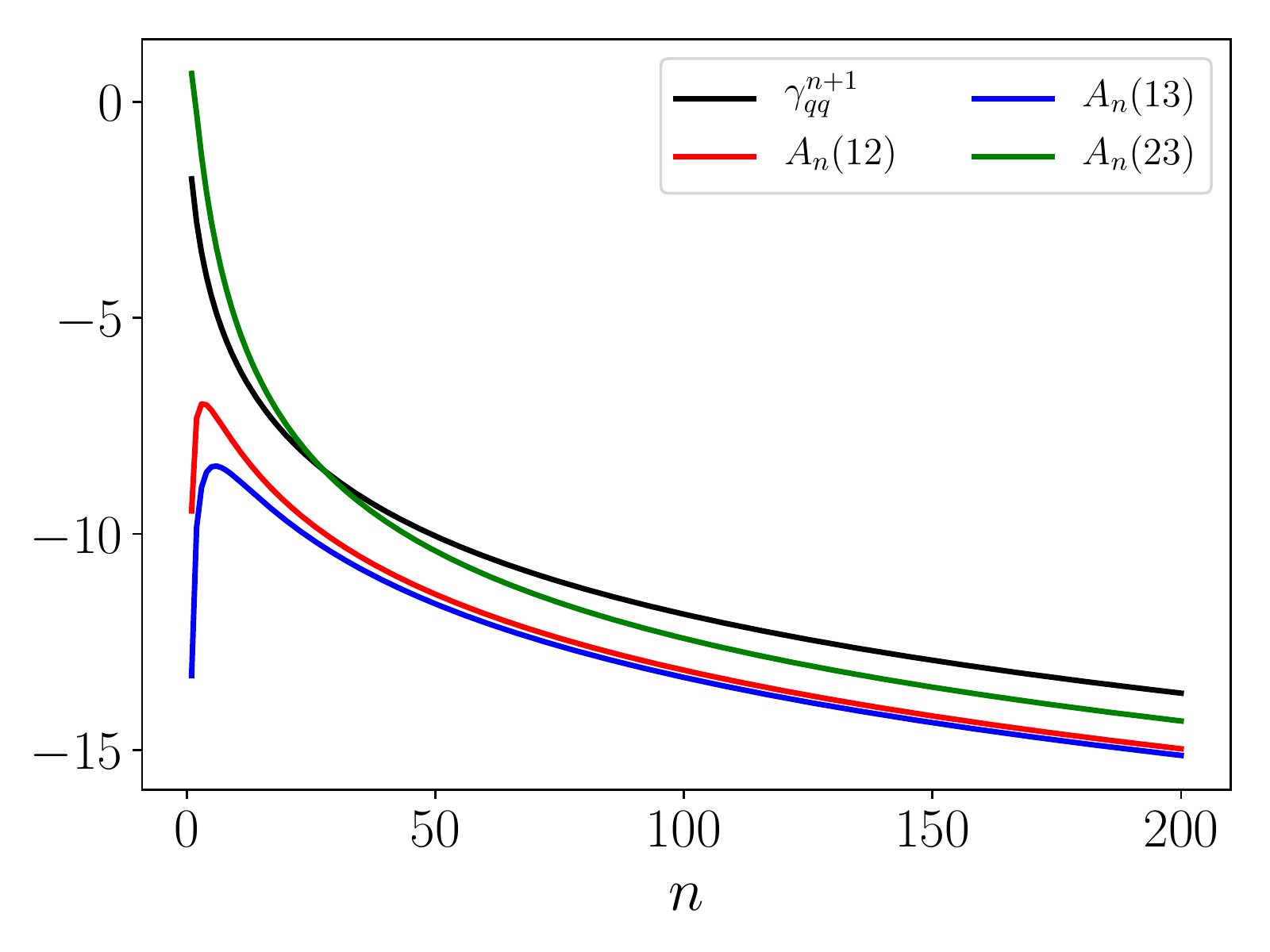}
        \caption{$A_n$'s and $\gamma^{n+1}_{qq}$.}
    \end{subfigure}
    ~ 
    \begin{subfigure}[t]{0.48\textwidth}
        \centering
        \includegraphics[height=2.2in]{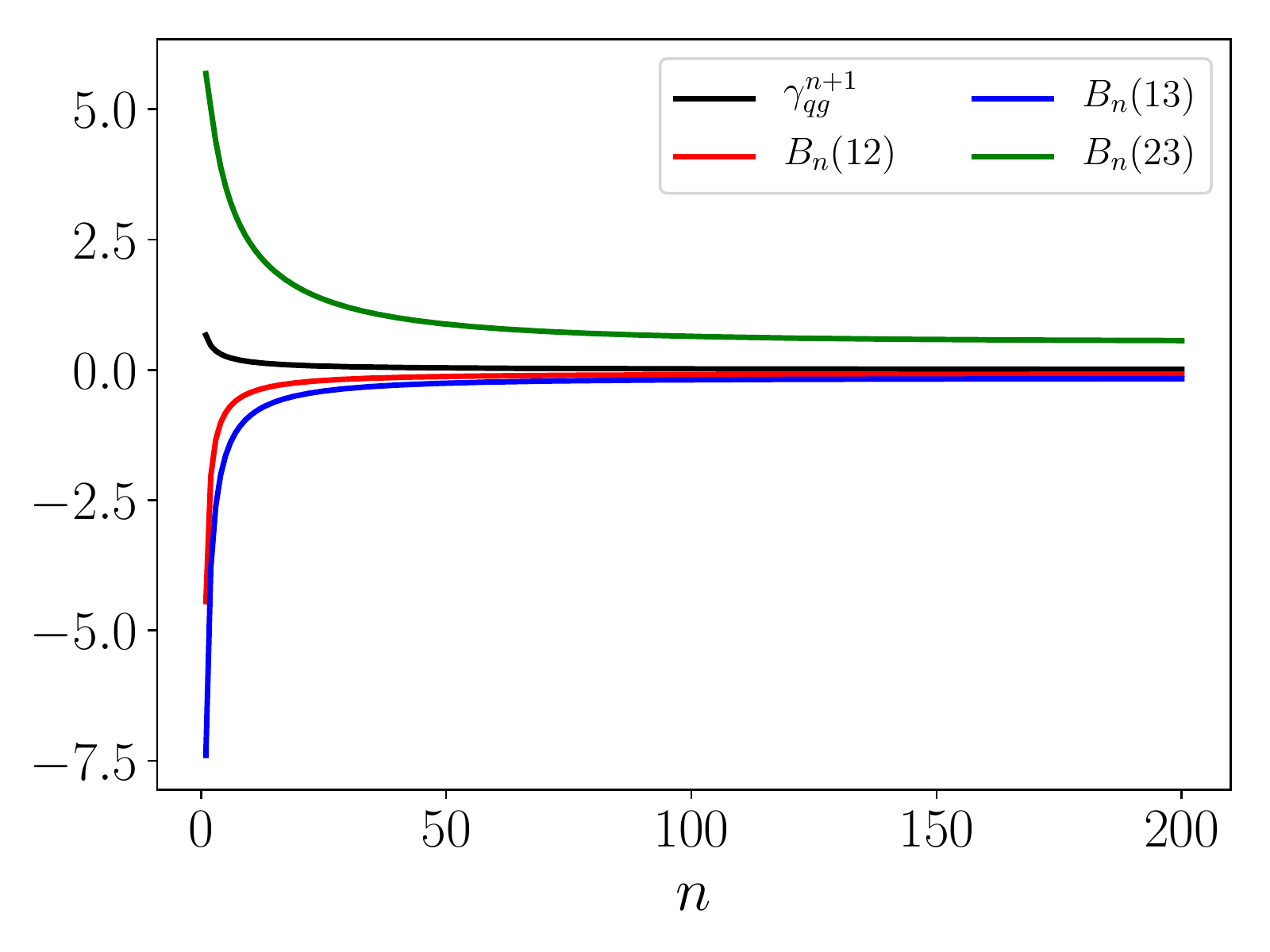}
        \caption{$B_n$'s and $\gamma^{n+1}_{qg}$.}
    \end{subfigure}%
    
    \begin{subfigure}[t]{0.48\textwidth}
        \centering
        \includegraphics[height=2.2in]{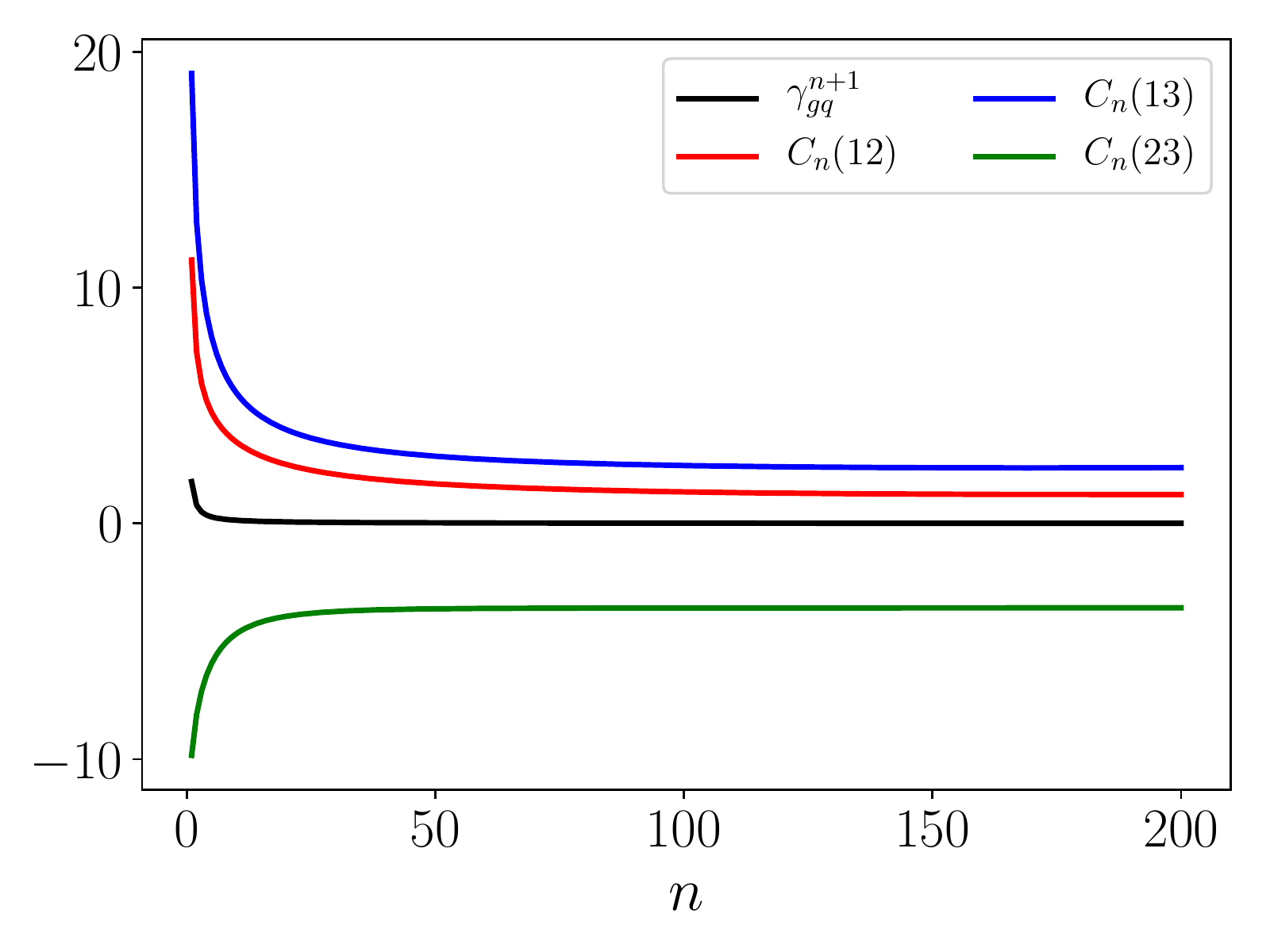}
        \caption{$C_n$'s and $\gamma^{n+1}_{gq}$.}
    \end{subfigure}%
    ~
    \begin{subfigure}[t]{0.48\textwidth}
        \centering
        \includegraphics[height=2.2in]{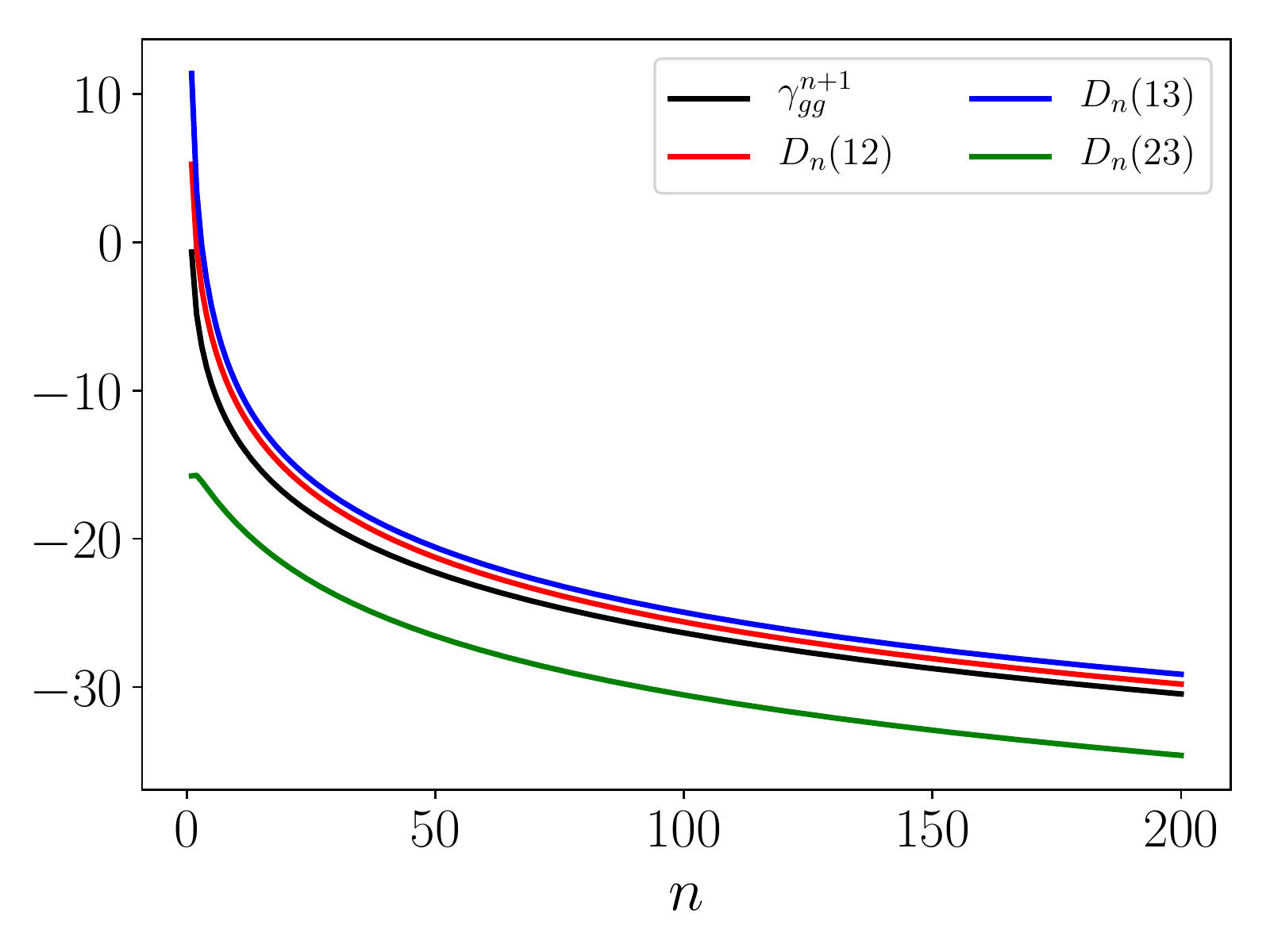}
        \caption{$D_n$'s and $\gamma^{n+1}_{gg}$.}
    \end{subfigure}%
    \caption{Comparison between the numerically extracted $A_n$, $B_n$, $C_n$ and $D_n$ and the anomalous dimensions of the twist-two operators.}
    \label{fig:anomalous}
\end{figure}

The comparison results are shown in Fig.~\ref{fig:anomalous}. The numerically extracted $A_n$, $B_n$, $C_n$ and $D_n$ of the twist-three operators are not universal because they depend on the choice of the initial condition. This implies that (\ref{naive}) is not valid and the moments of different $n$'s will mix with each other during the evolution. %The evolution equation for the twist-three operators is not closed itself. 
Interestingly, however, we find that the large-$n$ behaviors of the $A_n$, $B_n$, $C_n$ and $D_n$ are given by
\beq
A_n &=& -\frac{N_c^2-1}{N_c}\ln{n} + \ma{const.}  \label{co}  \\
B_n &=& \ma{const.} \nn
C_n &=& \ma{const.} \nn
D_n &=& -2N_c\ln{n}+ \ma{const.} \nonumber \,, 
\eeq
which agree with the $n$-dependence of $\gamma^{n+1}$'s when $n$ is large.  The constant terms in the $A_n$, $B_n$, $C_n$ and $D_n$ are not universal and depend on the choice of the initial conditions. The leading, logarithmic terms are called the cusp anomalous dimensions. 
We have thus found that, when $n$ is large, (\ref{naive}) is approximately valid and the second line of (\ref{naive2}) is small. This means that when $x\sim 1$, the total OAM distribution approximately satisfies the same evolution equation as in the WW approximation. We note that the emergence of the cusp anomalous dimension in the twist-three distributions in the large-$x$ limit was pointed out in \cite{Braun:2009mi} in a different context. Incidentally, we found the correct asymptotic behavior   (\ref{co}) only after fixing a typo in the code mentioned in Footnote 5 (the factor of $N_c$).
 %evolves with the universal cusp $n$-dependence of the $A_n$, $B_n$, $C_n$ and $D_n$ is universal when $n$ is large and is given by the cusp anomalous dimension. 
%However, in general (\ref{naive}) does not hold and moments with different $n$'s mix. This suggests that it is move convenient to study the evolution in $x$-space than in $n$-space.

\begin{figure}[h]
    \centering
    \begin{subfigure}[t]{0.33\textwidth}
        \centering
        \includegraphics[height=1.5in]{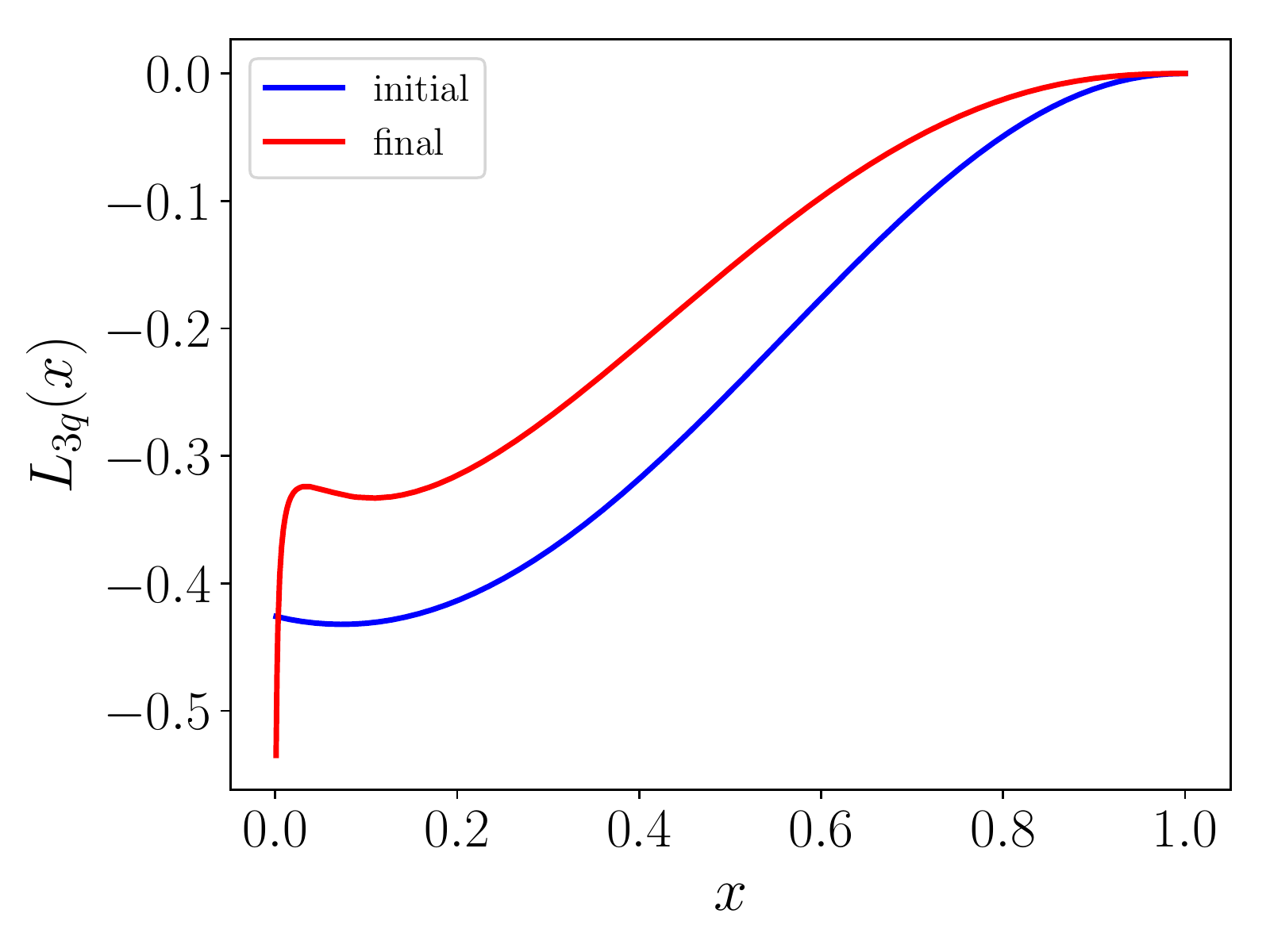}
        \caption{$L_{3q}(x)$ from initial condition 1.}
    \end{subfigure}%
    ~ 
    \begin{subfigure}[t]{0.33\textwidth}
        \centering
        \includegraphics[height=1.5in]{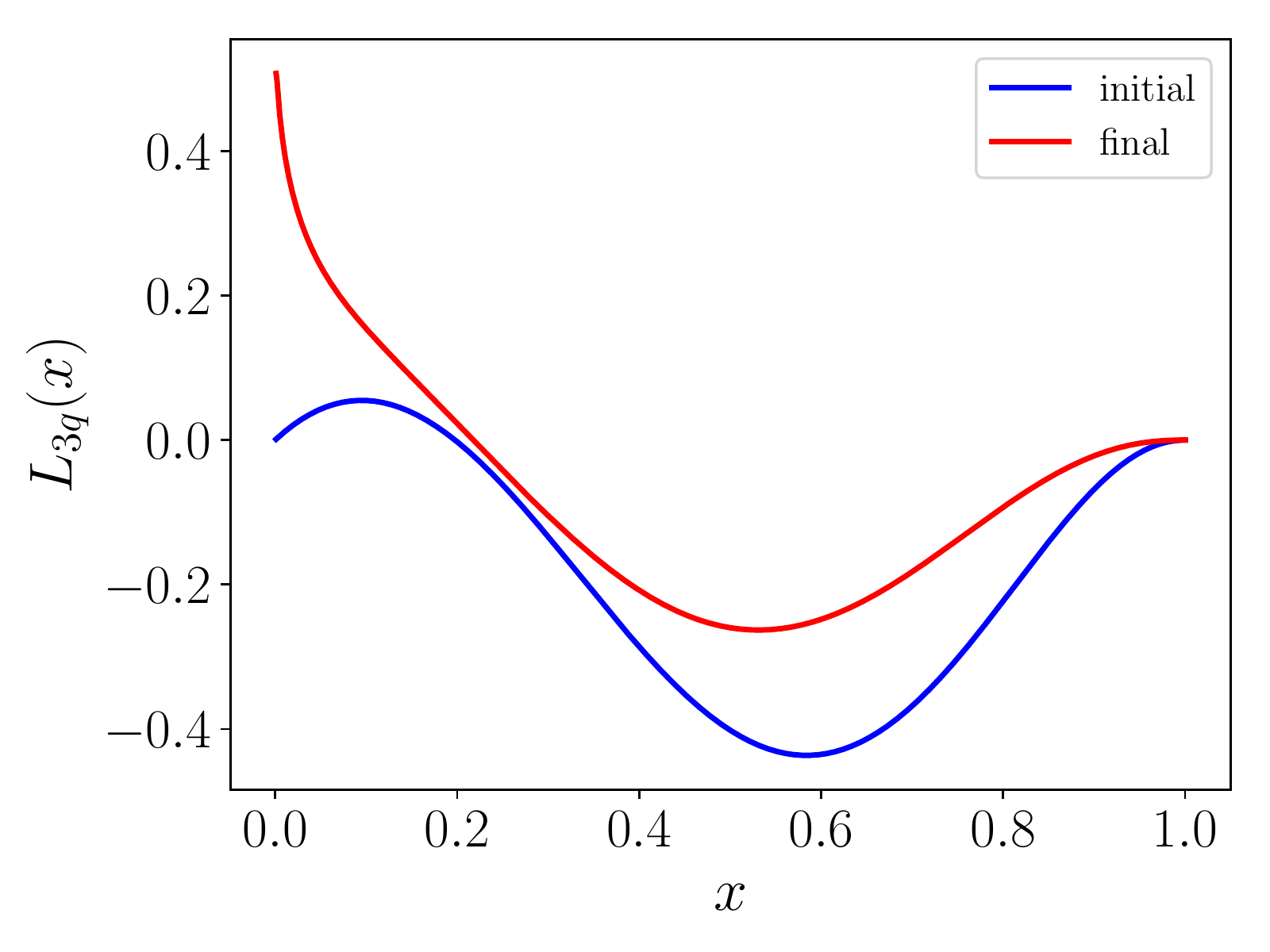}
        \caption{$L_{3q}(x)$ from initial condition 2.}
    \end{subfigure}%
    ~ 
    \begin{subfigure}[t]{0.33\textwidth}
        \centering
        \includegraphics[height=1.5in]{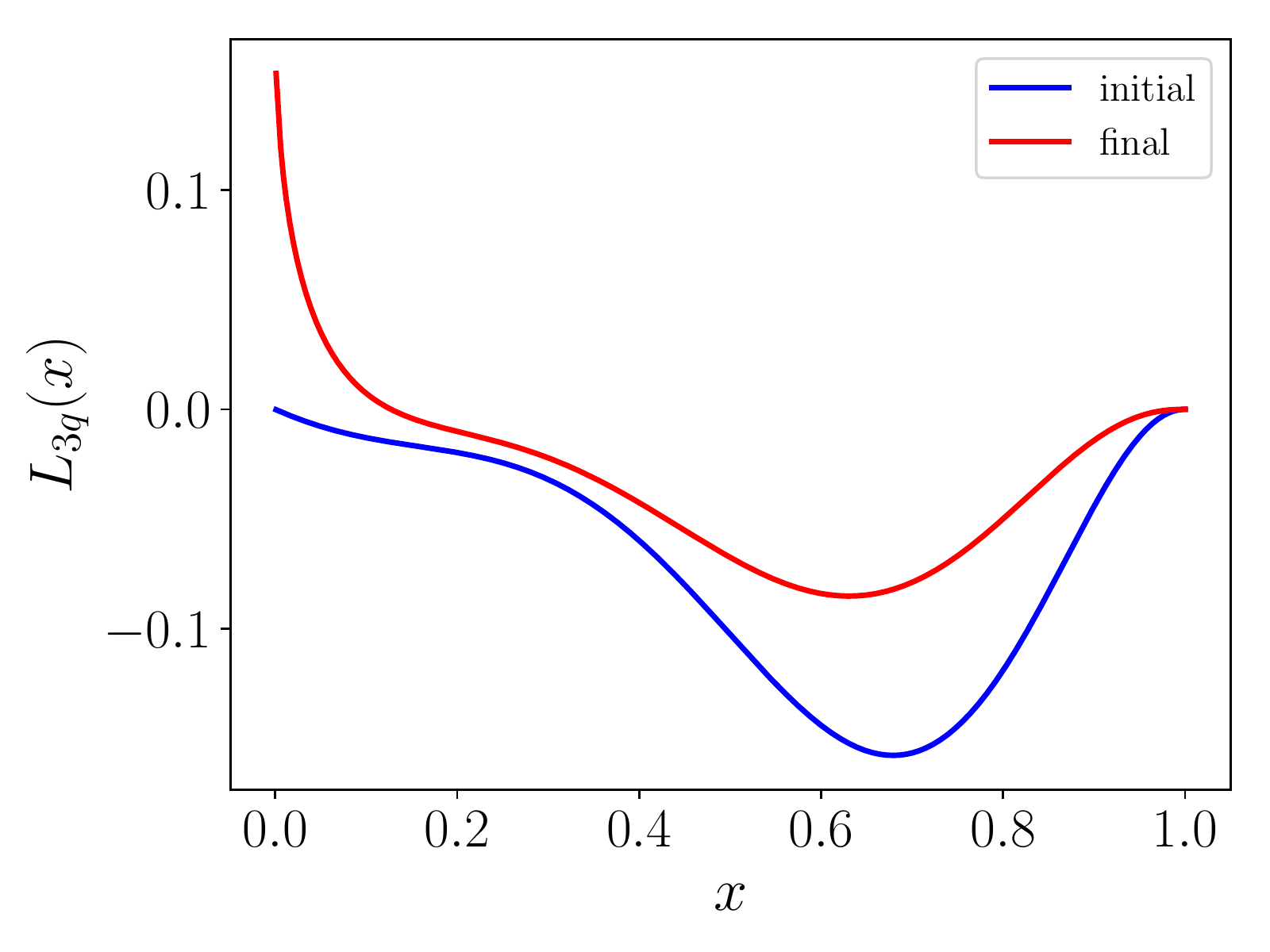}
        \caption{$L_{3q}(x)$ from initial condition 3.}
    \end{subfigure}%
    
    \begin{subfigure}[t]{0.33\textwidth}
        \centering
        \includegraphics[height=1.5in]{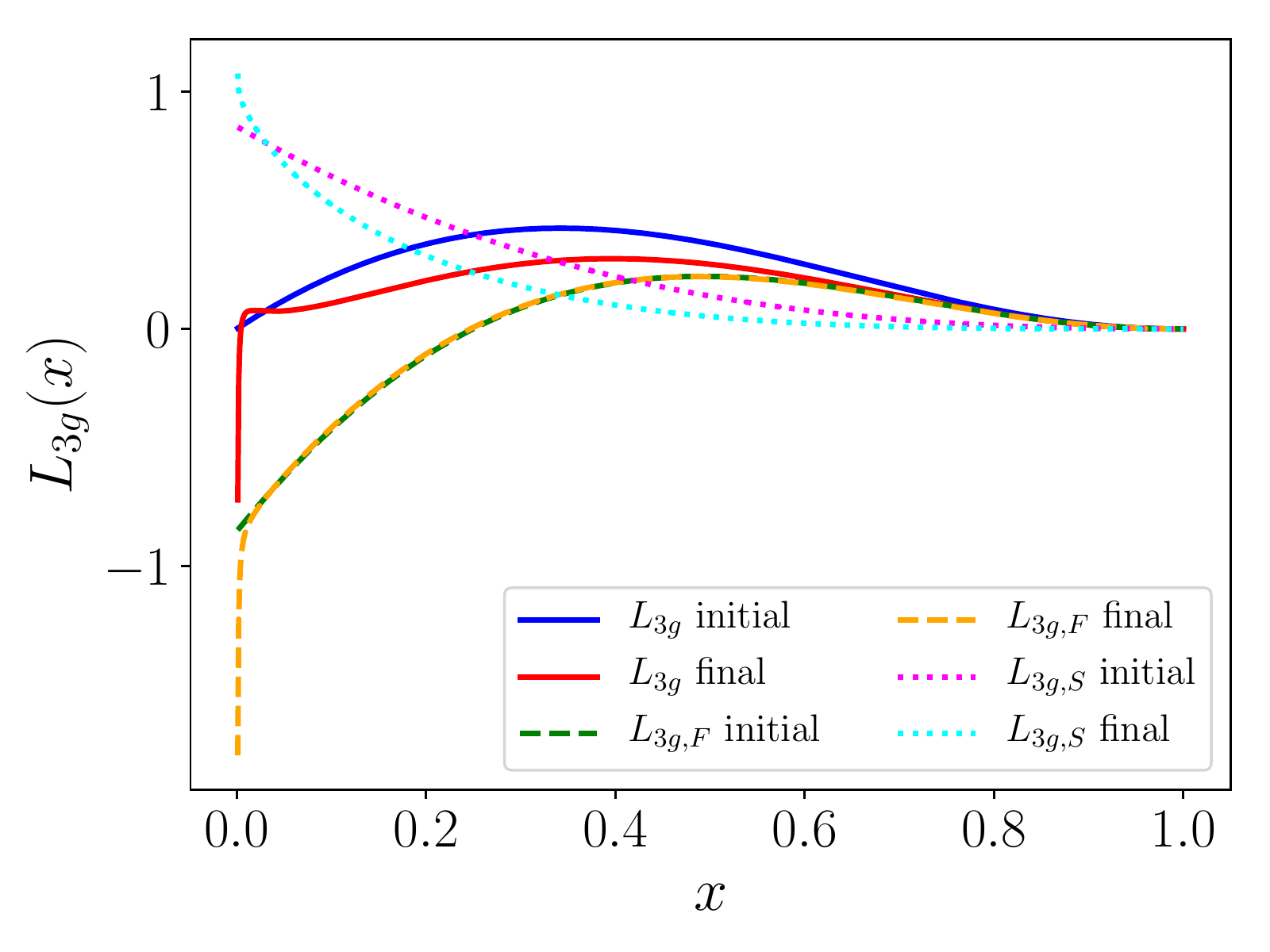}
        \caption{$L_{3g}(x)$ from initial condition 1.}
    \end{subfigure}%
    ~
    \begin{subfigure}[t]{0.33\textwidth}
        \centering
        \includegraphics[height=1.5in]{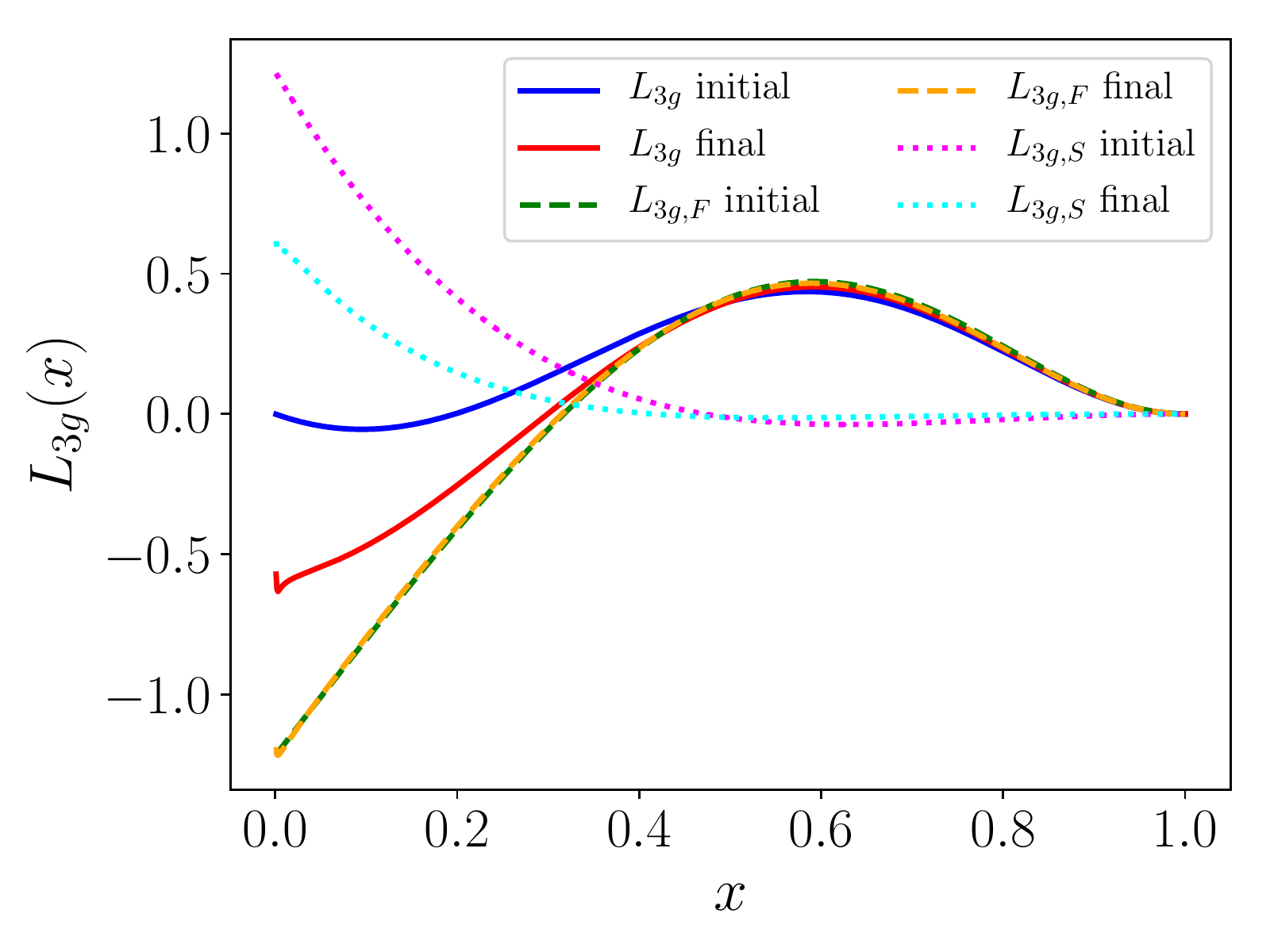}
        \caption{$L_{3g}(x)$ from initial condition 2.}
    \end{subfigure}%
    ~ 
    \begin{subfigure}[t]{0.33\textwidth}
        \centering
        \includegraphics[height=1.5in]{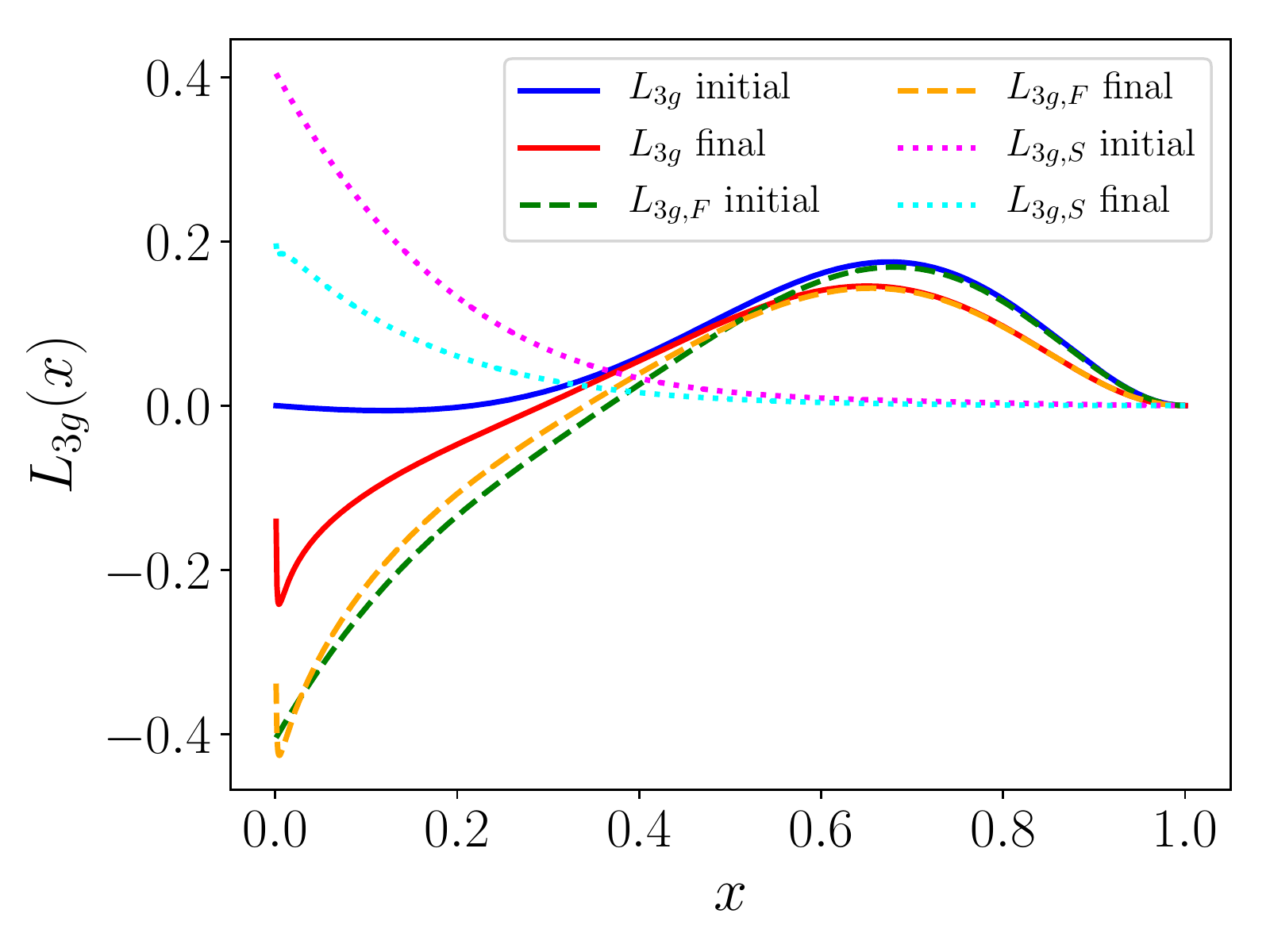}
        \caption{$L_{3g}(x)$ from initial condition 3.}
    \end{subfigure}%    
    \caption{Evolution of $L_{3q}(x)$ and $L_{3g}(x)$ from the initial $\mu_0^2=1$ GeV$^2$ to the final $\mu^2=10$ GeV$^2$ scales for three different initial conditions.}
    \label{fig:L}
\end{figure}

\begin{figure}[h]
\centering
\includegraphics[height=2.5in]{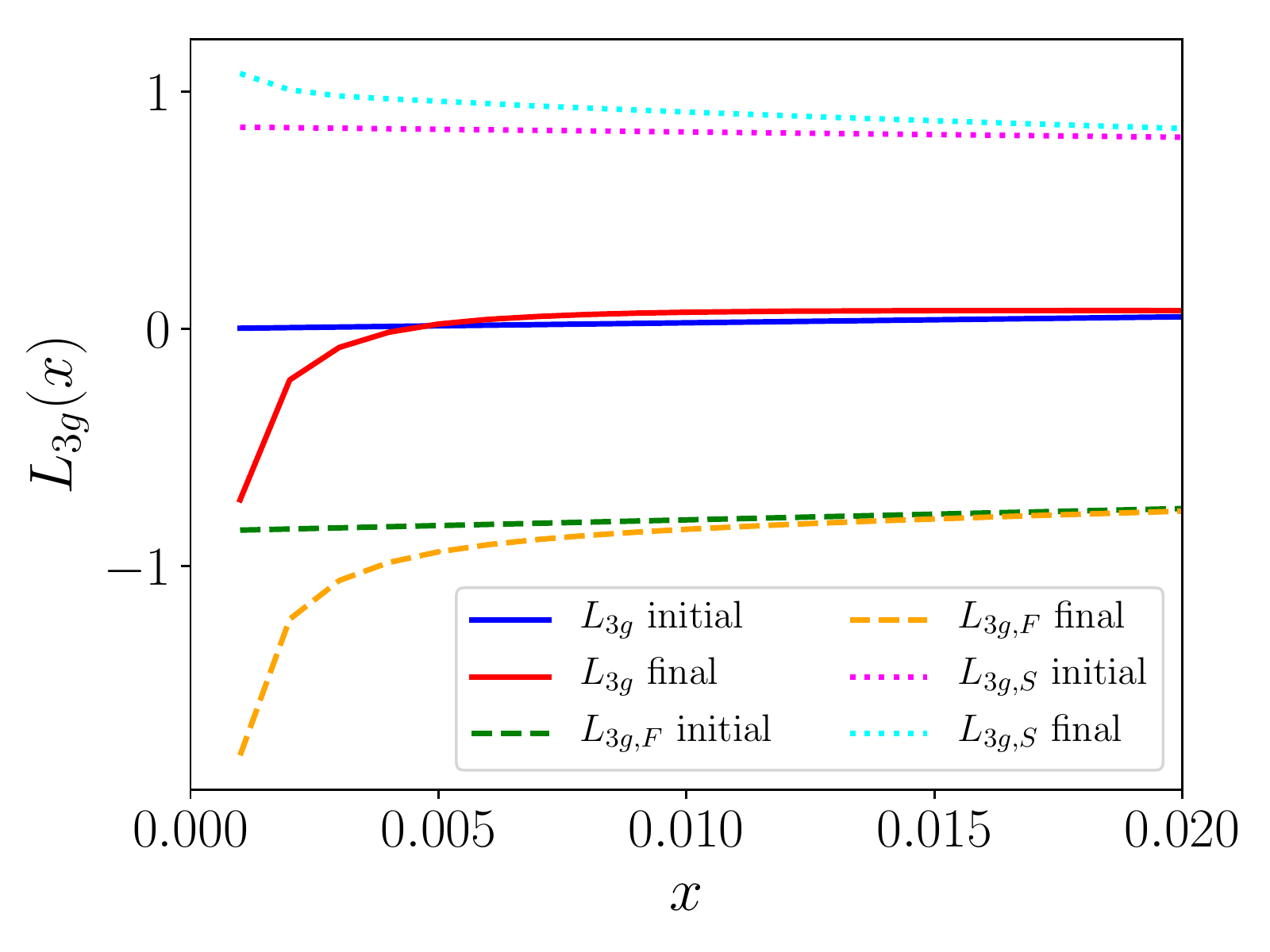}
\caption{Fig.~\ref{fig:L}(d), zoomed in on the small-$x$ region.}
\label{fig:L_3g}
\end{figure}

We then study the evolution from $\mu_0^2=1$ GeV$^2$ to $\mu^2=10$ GeV$^2$. The initial and final twist-three OAM distributions are plotted in Fig.~\ref{fig:L}. We note that both $L_{3q}(x)$ (flavor singlet) and $L_{3g}(x)$ develop singular behaviors at small-$x$ during the evolution, even though their initial conditions are regular at $x=0$. We fit the singular behavior by the power function $a x^b$ with $b<0$ and find that $b$ depends on the choice of initial conditions. But it always satisfies $b > -1$ in our three sets of initial conditions, which guarantees that both $L_{3q}(x)$ and $L_{3g}(x)$ are integrable in the range $0\leq x \leq 1$. The situation is somewhat similar to the DGLAP evolution of the WW part \cite{Hatta:2018itc}, where a singularity is developed from nonsingular initial conditions. In that case it was possible to gain some analytical insights, but in the present case a similar analytical study is difficult due to the complexity of the twist-three evolution. We also note that, curiously, the three-gluon part $L_{3g,F}(x)$ of $L_{3g}(x)$, see (\ref{d}), evolves very weakly with the renormalization scale. Since this part does not contribute to the integrated OAM (\ref{eqn:sum1}), we suspect that in general it plays a minor role in the nucleon spin decomposition. We however note that only for the initial condition 1, $L_{3g,F}(x)$ shows a sharp drop in the smallest $x$ bins that can be fitted by a power law as shown in Fig.~\ref{fig:L_3g}. In fact, such a rapid behavior is necessary to satisfy the sum rule in this case. 

Finally we study the $\mu^2$-dependence of the potential  angular momentum (\ref{pot}). Here we are also interested in the flavor nonsinglet part $L_\ma{pot}^{u-d}$ because  the recent lattice calculations have shown that this quantity is positive and large \cite{Engelhardt:2017miy,Engelhardt:2019lyy}. In fact, the singlet part $L^{u+d}_\ma{pot}$ is numerically much smaller due to  a large cancelation between the $u$- and $d$-quark contributions.  We have not mentioned the evolution in the flavor nonsinglet sector so far, but of course it is simpler than  the singlet case because there is  no mixing with the three-gluon part. The code can handle this as well.  

\begin{figure}[h]
    \centering
    \begin{subfigure}[t]{0.48\textwidth}
        \centering
        \includegraphics[height=2.5in]{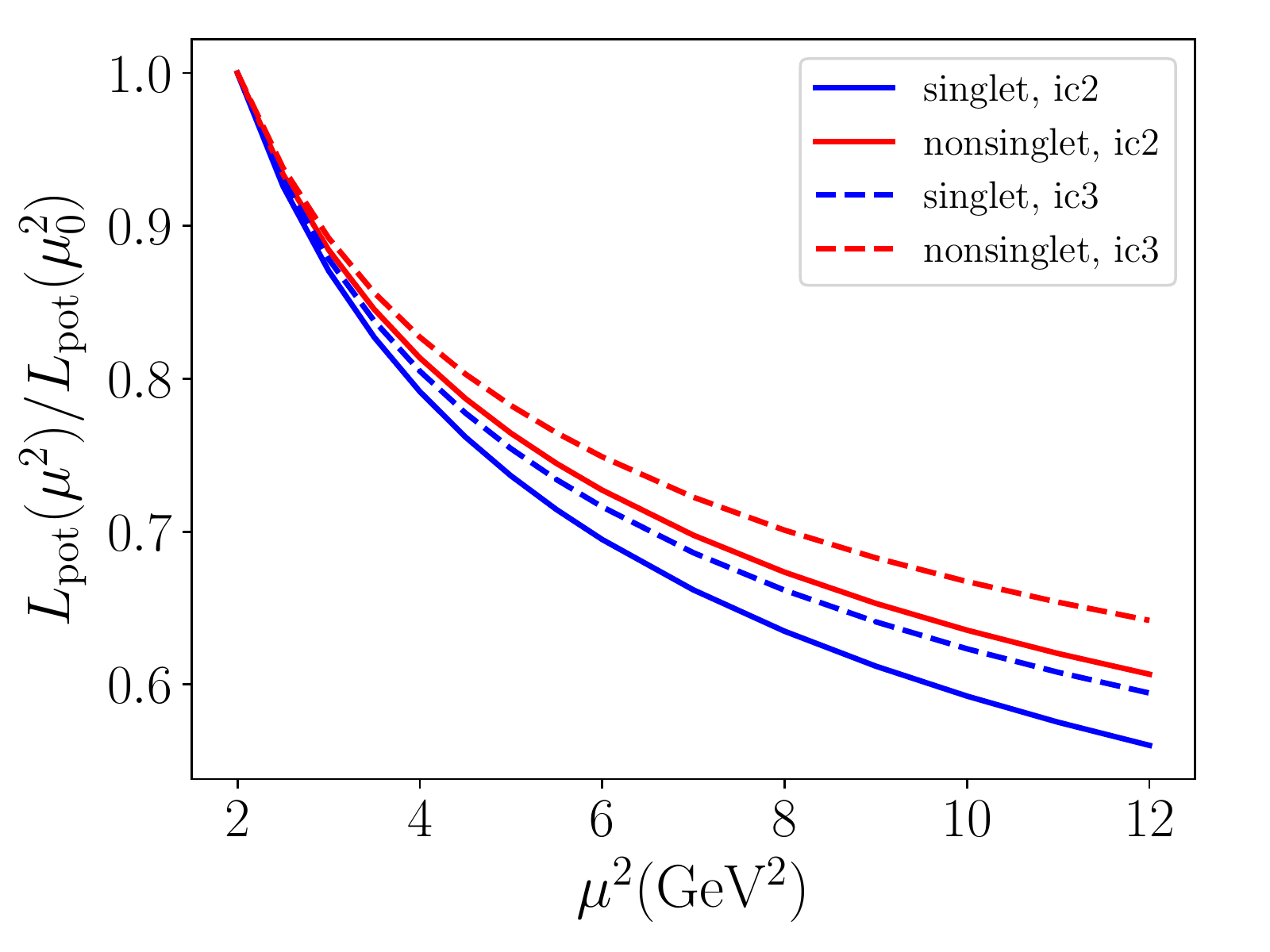}
        \caption{Scale dependence of $L_\ma{pot}$.}
    \end{subfigure}
    ~ 
    \begin{subfigure}[t]{0.48\textwidth}
        \centering
        \includegraphics[height=2.5in]{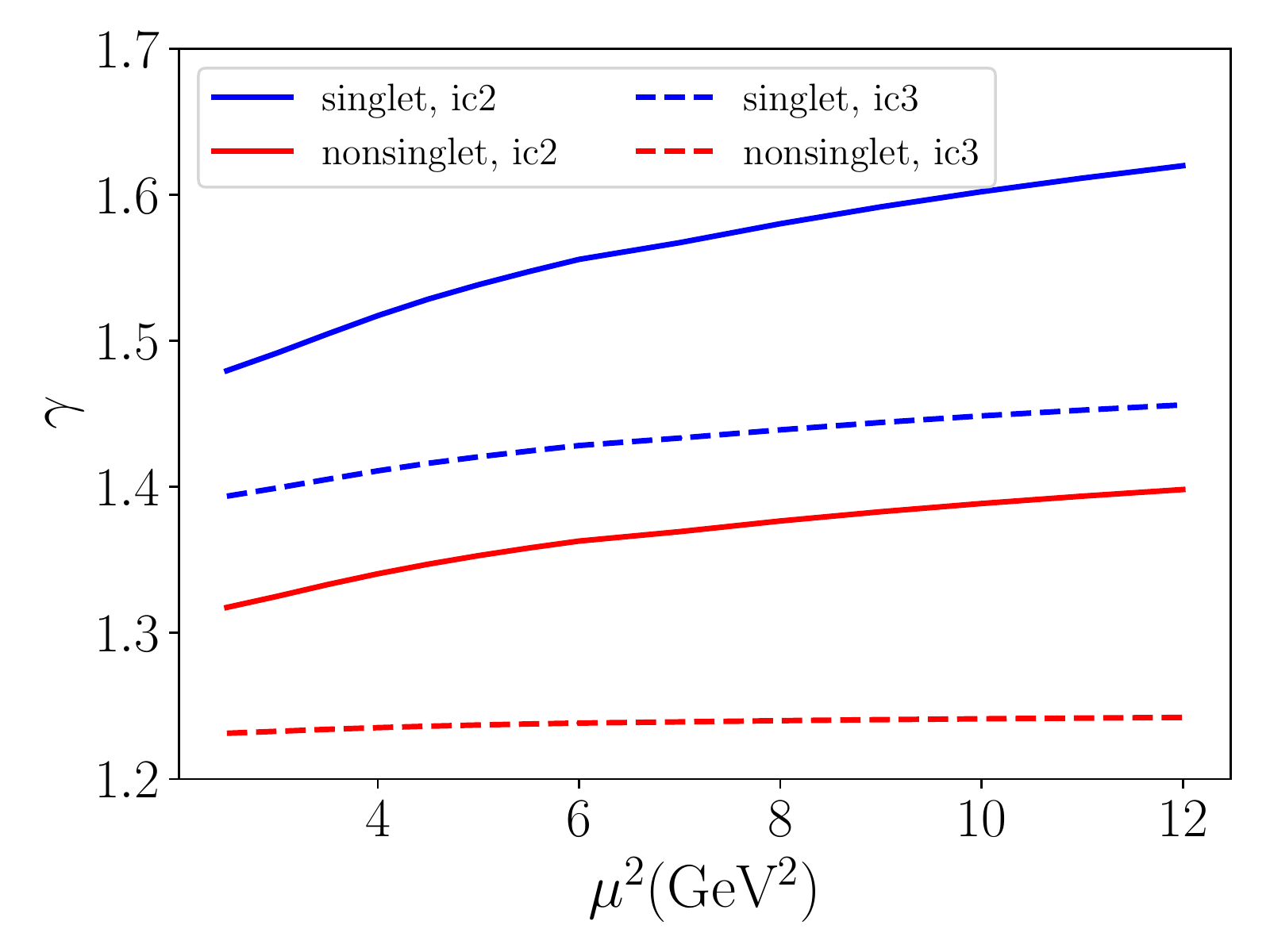}
        \caption{Scale dependence of $\gamma$ in (\ref{ga}).}
    \end{subfigure}%
\caption{Evolution of $L_\ma{pot}$ for initial conditions 2 and 3.}
\label{fig:L_pot}
\end{figure}

%We assume $L_\ma{pot}^{singlet}>0$ and $L_\ma{pot}^{u-d}>0$ as suggested by the results in \cite{Engelhardt:2017miy}. 
%\beq
%L_\ma{pot} = \int_0^1 dx L_{3\Sigma}(x)\,,
%\eeq
%which is the difference between the quark OAM defined by Jaffe-Manohar and that defined by Ji:
%\beq
%L_\ma{pot} = L_\Sigma^{JM} - L_\Sigma^{Ji} \,.
%\eeq
We evolve the initial conditions 2 and 3 from $\mu_0^2=2$ GeV$^2$ to $\mu^2=12$ GeV$^2$ and compute how $L_\ma{pot}$ changes with the scale for both the flavor singlet and nonsinglet parts. The initial and final scales are chosen such that the number of dynamical quarks $n_f$ in the running coupling does not change in the course of evolution. The results are shown in the left panel of Fig.~\ref{fig:L_pot}. In both cases, the evolution tends to suppress the magnitude of $L_\ma{pot}$, with a stronger suppression in the singlet case. This can be understood as arising from the mixing with the three-gluon correlator. 
We have fitted these results in the form 
%Approximately, the evolution of $L_\ma{pot}$ for both the flavor singlet and nonsinglet can be described by
\beq
L_\ma{pot}(\mu^2) = L_\ma{pot}(\mu_0^2) \bigg(   \frac{ \ln{\mu_0^2/\Lambda^2_{QCD}} }{ \ln{\mu^2/\Lambda^2_{QCD}} }   \bigg)^\gamma \,, \label{ga}
\eeq
which features an  `anomalous dimension' $\gamma$. Here $\Lambda_{QCD}=0.175$ GeV is used in the code for $n_f=4$. We find that $\gamma$ is $\mu$-dependent as shown in the right panel of Fig.~\ref{fig:L_pot}. It also depends on the initial condition. This means that the evolution of $L_\ma{pot}$ cannot be characterized by a single anomalous dimension even in the nonsinglet sector due to the mixing between different moments.
%where $\gamma$ should not be thought of as the anomalous dimension because it is not universal and depends on the scale. The scale dependence of $\gamma$ is also plotted in Fig.~\ref{fig:L_pot}. 
From the obtained behavior of $\gamma$, we deduce that $L_\ma{pot}\to 0$ as $\mu \to \infty$ (though this has to be checked more carefully). Therefore, the asymptotic result for the integrated OAM
\beq
L_q(\mu\to \infty) \approx  -\frac{1}{2}\Delta \Sigma  +\frac{1}{2}\frac{3n_f}{16+3n_f}, \qquad L_g(\mu\to \infty) \approx -\Delta G(\mu) +\frac{1}{2}\frac{16}{16+3n_f},
\label{sub}
\eeq
obtained in the WW approximation \cite{Ji:1995cu}  will not be affected. The subleading corrections to (\ref{sub}) have the $\mu$-dependence of the form (\ref{ga}) with $\gamma= \frac{2(16+3n_f)}{9b_0} \approx 0.75$ (for $n_f=4$) \cite{Ji:1995cu}. Since this is smaller than the value shown in Fig.~\ref{fig:L_pot}, it seems that $L_\ma{pot}$ can be  neglected in the large-$\mu$ region.

In \cite{Ji:2015sio}, it has been shown that $L_\ma{pot}$ vanishes for a single electron state to one-loop order in QED. This is not  necessarily in contradiction to the present result. The operator 
$\vec{x}\times \bar{\psi}\gamma^+  \vec{A}\psi$ (in the light-cone gauge) relevant to $L_\ma{pot}$ mixes with a tower of operators $\vec{x}\times\bar{\psi}\gamma^+  D^\mu D^\nu\cdots \vec{A}\psi $ and if these higher moments have  nonvanishing initial values (as in our initial conditions), they will affect the renormalization of $L_\ma{pot}$. 
  
\section{Conclusions}
\label{sec5}
In this paper, we have studied, for the first time, the one-loop QCD evolution of the genuine twist-three part of the OAM distributions. In particular, the scale variation of the potential angular momentum has been demonstrated. As anticipated by the complicated relations between $L_{3q,3g}(x)$ and the underlying distributions $S^+$ and $F^+$ (see, e.g., (\ref{final1})), different moments of $L_{3q,3g}(x)$ mix under evolution except in the large-$n$ limit. This suggests that it is more convenient to look at the evolution directly in the $x$-space.       
Together with the known evolution of the WW part, the one-loop evolution of the total OAM distributions $L_{q,g}(x)$ is now fully under control and ready for phenomenological applications. 

In the present C++ code, the grid in $x$ is uniform  and the smallest value of $x$ that we achieved is 0.001. It is not realistic to go to much lower values of $x$ because it is computationally too expensive. In view of the recent controversy  regarding the small-$x$ asymptotic behavior of the OAM distributions \cite{Hatta:2018itc,Kovchegov:2019rrz,Boussarie:2019icw}, it would be interesting to modify the code (e.g., set up a grid in $\ln 1/x$ instead of $x$) to zoom in on to the small-$x$ region as was done for the WW part \cite{Hatta:2018itc}.  We leave this to future work.

\section*{Acknowledgments}

We thank Renaud Boussarie, Vladimir Braun,  Alexander Manashov and Werner Vogelsang for discussions. This material is based upon work supported by  the U.S. Department of Energy, Office of Science, Office of Nuclear Physics, under contract number  DE-SC0012704. It is also supported by the LDRD program of Brookhaven National Laboratory. X.Y. is supported by U.S. Department of Energy research grant DE-FG02-05ER41367 and Brookhaven National Laboratory.


\begin{thebibliography}{99}

%\cite{Adare:2008aa}
\bibitem{Adare:2008aa} 
  A.~Adare {\it et al.} [PHENIX Collaboration],
  %``The Polarized gluon contribution to the proton spin from the double helicity asymmetry in inclusive pi0 production in polarized p + p collisions at s**(1/2) = 200-GeV,''
  Phys.\ Rev.\ Lett.\  {\bf 103}, 012003 (2009)
  %doi:10.1103/PhysRevLett.103.012003
  [arXiv:0810.0694 [hep-ex]].
  %%CITATION = doi:10.1103/PhysRevLett.103.012003;%%
  %111 citations counted in INSPIRE as of 17 Jun 2019
  

  %\cite{Airapetian:2010ac}
\bibitem{Airapetian:2010ac} 
  A.~Airapetian {\it et al.} [HERMES Collaboration],
  %``Leading-Order Determination of the Gluon Polarization from high-p(T) Hadron Electroproduction,''
  JHEP {\bf 1008}, 130 (2010)
  %doi:10.1007/JHEP08(2010)130
  [arXiv:1002.3921 [hep-ex]].
  %%CITATION = doi:10.1007/JHEP08(2010)130;%%
  %62 citations counted in INSPIRE as of 17 Jun 2019
  
  %\cite{Adamczyk:2012qj}
\bibitem{Adamczyk:2012qj} 
  L.~Adamczyk {\it et al.} [STAR Collaboration],
  %``Longitudinal and transverse spin asymmetries for inclusive jet production at mid-rapidity in polarized $p+p$ collisions at $\sqrt{s}=200$ GeV,''
  Phys.\ Rev.\ D {\bf 86}, 032006 (2012)
  %doi:10.1103/PhysRevD.86.032006
  [arXiv:1205.2735 [nucl-ex]].
  %%CITATION = doi:10.1103/PhysRevD.86.032006;%%
  %80 citations counted in INSPIRE as of 17 Jun 2019

  %\cite{Alekseev:2010ub}
\bibitem{Alekseev:2010ub} 
  M.~G.~Alekseev {\it et al.} [COMPASS Collaboration],
  %``Quark helicity distributions from longitudinal spin asymmetries in muon-proton and muon-deuteron scattering,''
  Phys.\ Lett.\ B {\bf 693}, 227 (2010)
  %doi:10.1016/j.physletb.2010.08.034
  [arXiv:1007.4061 [hep-ex]].
  %%CITATION = doi:10.1016/j.physletb.2010.08.034;%%
  %158 citations counted in INSPIRE as of 17 Jun 2019

%\cite{Adamczyk:2014ozi}
\bibitem{Adamczyk:2014ozi} 
  L.~Adamczyk {\it et al.} [STAR Collaboration],
  %``Precision Measurement of the Longitudinal Double-spin Asymmetry for Inclusive Jet Production in Polarized Proton Collisions at $\sqrt{s}=200$ GeV,''
  Phys.\ Rev.\ Lett.\  {\bf 115}, no. 9, 092002 (2015)
  %doi:10.1103/PhysRevLett.115.092002
  [arXiv:1405.5134 [hep-ex]].
  %%CITATION = doi:10.1103/PhysRevLett.115.092002;%%
  %96 citations counted in INSPIRE as of 17 Jun 2019


%\cite{Prok:2014ltt}
\bibitem{Prok:2014ltt} 
  Y.~Prok {\it et al.} [CLAS Collaboration],
  %``Precision measurements of $g_1$ of the proton and the deuteron with 6 GeV electrons,''
  Phys.\ Rev.\ C {\bf 90}, no. 2, 025212 (2014)
  %doi:10.1103/PhysRevC.90.025212
  [arXiv:1404.6231 [nucl-ex]].
  %%CITATION = doi:10.1103/PhysRevC.90.025212;%%
  %26 citations counted in INSPIRE as of 17 Jun 2019

%\cite{deFlorian:2014yva}
\bibitem{deFlorian:2014yva} 
  D.~de Florian, R.~Sassot, M.~Stratmann and W.~Vogelsang,
  %``Evidence for polarization of gluons in the proton,''
  Phys.\ Rev.\ Lett.\  {\bf 113}, no. 1, 012001 (2014)
  %doi:10.1103/PhysRevLett.113.012001
  [arXiv:1404.4293 [hep-ph]].
  %%CITATION = doi:10.1103/PhysRevLett.113.012001;%%
  %200 citations counted in INSPIRE as of


%\cite{Nocera:2014gqa}
\bibitem{Nocera:2014gqa} 
  E.~R.~Nocera {\it et al.} [NNPDF Collaboration],
  %``A first unbiased global determination of polarized PDFs and their uncertainties,''
  Nucl.\ Phys.\ B {\bf 887}, 276 (2014)
  %doi:10.1016/j.nuclphysb.2014.08.008
  [arXiv:1406.5539 [hep-ph]].
  %%CITATION = doi:10.1016/j.nuclphysb.2014.08.008;%%
  %175 citations counted in INSPIRE as of 16 Jun


%\cite{Sato:2016tuz}
\bibitem{Sato:2016tuz} 
  N.~Sato {\it et al.} [Jefferson Lab Angular Momentum Collaboration],
  %``Iterative Monte Carlo analysis of spin-dependent parton distributions,''
  Phys.\ Rev.\ D {\bf 93}, no. 7, 074005 (2016)
  %doi:10.1103/PhysRevD.93.074005
  [arXiv:1601.07782 [hep-ph]].
  %%CITATION = doi:10.1103/PhysRevD.93.074005;%%
  %67 citations counted in INSPIRE

%\cite{Jaffe:1989jz}
\bibitem{Jaffe:1989jz} 
  R.~L.~Jaffe and A.~Manohar,
  %``The G(1) Problem: Fact and Fantasy on the Spin of the Proton,''
  Nucl.\ Phys.\ B {\bf 337}, 509 (1990).
  %doi:10.1016/0550-3213(90)90506-9
  %%CITATION = doi:10.1016/0550-3213(90)90506-9;%%
  %805 citations counted in INSPIRE as of


%\cite{Adam:2019aml}
\bibitem{Adam:2019aml} 
  J.~Adam {\it et al.} [STAR Collaboration],
  %``Longitudinal double-spin asymmetry for inclusive jet and dijet production in $pp$ collisions at $\sqrt{s} = 510$ GeV,''
  arXiv:1906.02740 [hep-ex].
  %%CITATION = ARXIV:1906.02740;%%

%\cite{Accardi:2012qut}
\bibitem{Accardi:2012qut} 
  A.~Accardi {\it et al.},
  %``Electron Ion Collider: The Next QCD Frontier : Understanding the glue that binds us all,''
  Eur.\ Phys.\ J.\ A {\bf 52}, no. 9, 268 (2016)
  %doi:10.1140/epja/i2016-16268-9
  [arXiv:1212.1701 [nucl-ex]].
  %%CITATION = doi:10.1140/epja/i2016-16268-9;%%
  %582 citations counted in INSPIRE as 



%\cite{Hatta:2012cs}
\bibitem{Hatta:2012cs} 
  Y.~Hatta and S.~Yoshida,
  %``Twist analysis of the nucleon spin in QCD,''
  JHEP {\bf 1210}, 080 (2012)
  %doi:10.1007/JHEP10(2012)080
  [arXiv:1207.5332 [hep-ph]].
  %%CITATION = doi:10.1007/JHEP10(2012)080;%%
  %55 citations counted

%\cite{Lorce:2011kd}
\bibitem{Lorce:2011kd} 
  C.~Lorce and B.~Pasquini,
  %``Quark Wigner Distributions and Orbital Angular Momentum,''
  Phys.\ Rev.\ D {\bf 84}, 014015 (2011)
  %doi:10.1103/PhysRevD.84.014015
  [arXiv:1106.0139 [hep-ph]].
  %%CITATION = doi:10.1103/PhysRevD.84.014015;%%
  %182 citations counted in INSPIRE as 

%\cite{Hatta:2011ku}
\bibitem{Hatta:2011ku} 
  Y.~Hatta,
  %``Notes on the orbital angular momentum of quarks in the nucleon,''
  Phys.\ Lett.\ B {\bf 708}, 186 (2012)
  %doi:10.1016/j.physletb.2012.01.024
  [arXiv:1111.3547 [hep-ph]].
  %%CITATION = doi:10.1016/j.physletb.2012.01.024;%%
  %136 citations counted in INSP



%\cite{Lorce:2011ni}
\bibitem{Lorce:2011ni} 
  C.~Lorce, B.~Pasquini, X.~Xiong and F.~Yuan,
  %``The quark orbital angular momentum from Wigner distributions and light-cone wave functions,''
  Phys.\ Rev.\ D {\bf 85}, 114006 (2012)
  %doi:10.1103/PhysRevD.85.114006
  [arXiv:1111.4827 [hep-ph]].
  %%CITATION = doi:10.1103/PhysRevD.85.114006;%%
  %113 citations counted in INSPIRE as





%\cite{Harindranath:1998ve}
\bibitem{Harindranath:1998ve} 
  A.~Harindranath and R.~Kundu,
  %``On Orbital angular momentum in deep inelastic scattering,''
  Phys.\ Rev.\ D {\bf 59}, 116013 (1999)
  %doi:10.1103/PhysRevD.59.116013
  [hep-ph/9802406].
  %%CITATION = doi:10.1103/PhysRevD.59.116013;%%
  %88 citations counted in INSPIRE as

%\cite{Hagler:1998kg}
\bibitem{Hagler:1998kg} 
  P.~Hagler and A.~Schafer,
  %``Evolution equations for higher moments of angular momentum distributions,''
  Phys.\ Lett.\ B {\bf 430}, 179 (1998)
  %doi:10.1016/S0370-2693(98)00414-6
  [hep-ph/9802362].
  %%CITATION = doi:10.1016/S0370-2693(98)00414-6;%%
  %44 citations counted in INSPIRE

  
  %\cite{Courtoy:2013oaa}
\bibitem{Courtoy:2013oaa} 
  A.~Courtoy, G.~R.~Goldstein, J.~O.~Gonzalez Hernandez, S.~Liuti and A.~Rajan,
  %``On the Observability of the Quark Orbital Angular Momentum Distribution,''
  Phys.\ Lett.\ B {\bf 731}, 141 (2014)
  %doi:10.1016/j.physletb.2014.02.017
  [arXiv:1310.5157 [hep-ph]].
  %%CITATION = doi:10.1016/j.physletb.2014.02.017;%%
  %38 citations counted in INSPIRE as of 17 Jun 2019
  
%\cite{Ji:2016jgn}
\bibitem{Ji:2016jgn} 
  X.~Ji, F.~Yuan and Y.~Zhao,
  %``Hunting the Gluon Orbital Angular Momentum at the Electron-Ion Collider,''
  Phys.\ Rev.\ Lett.\  {\bf 118}, no. 19, 192004 (2017)
  %doi:10.1103/PhysRevLett.118.192004
  [arXiv:1612.02438 [hep-ph]].
  %%CITATION = doi:10.1103/PhysRevLett.118.192004;%%
  %16 citations counted in INSPIRE as of 


%\cite{Hatta:2016aoc}
\bibitem{Hatta:2016aoc} 
  Y.~Hatta, Y.~Nakagawa, F.~Yuan, Y.~Zhao and B.~Xiao,
  %``Gluon orbital angular momentum at small-$x$,''
  Phys.\ Rev.\ D {\bf 95}, no. 11, 114032 (2017)
  %doi:10.1103/PhysRevD.95.114032
  [arXiv:1612.02445 [hep-ph]].
  %%CITATION = doi:10.1103/PhysRevD.95.114032;%%
  %22 citations counted in INSPIRE as of 07


%\cite{Bhattacharya:2017bvs}
\bibitem{Bhattacharya:2017bvs} 
  S.~Bhattacharya, A.~Metz and J.~Zhou,
  %``Generalized TMDs and the exclusive double Drell–Yan process,''
  Phys.\ Lett.\ B {\bf 771}, 396 (2017)
  %doi:10.1016/j.physletb.2017.05.081
  [arXiv:1702.04387 [hep-ph]].
  %%CITATION = doi:10.1016/j.physletb.2017.05.081;%%
  %24 citations counted in INSP

%\cite{Bhattacharya:2018lgm}
\bibitem{Bhattacharya:2018lgm} 
  S.~Bhattacharya, A.~Metz, V.~K.~Ojha, J.~Y.~Tsai and J.~Zhou,
  %``Exclusive double quarkonium production and generalized TMDs of gluons,''
  arXiv:1802.10550 [hep-ph].
  %%CITATION = ARXIV:1802.10550;%%
  %4 citations counted in INSPIRE as 





%\cite{Hoodbhoy:1998yb}
\bibitem{Hoodbhoy:1998yb} 
  P.~Hoodbhoy, X.~D.~Ji and W.~Lu,
  %``Quark orbital - angular - momentum distribution in the nucleon,''
  Phys.\ Rev.\ D {\bf 59}, 014013 (1999)
  %doi:10.1103/PhysRevD.59.014013
  [hep-ph/9804337].
  %%CITATION = doi:10.1103/PhysRevD.59.014013;%%
  %68 citations counted in INSPIR

%\cite{Boussarie:2019icw}
\bibitem{Boussarie:2019icw} 
  R.~Boussarie, Y.~Hatta and F.~Yuan,
  %``Proton Spin Structure at Small-$x$,''
  arXiv:1904.02693 [hep-ph].
  %%CITATION = ARXIV:1904.02693;%%






%\cite{Hatta:2018itc}
\bibitem{Hatta:2018itc} 
  Y.~Hatta and D.~J.~Yang,
  %``On the small-$x$ behavior of the orbital angular momentum distributions in QCD,''
  Phys.\ Lett.\ B {\bf 781}, 213 (2018)
  %doi:10.1016/j.physletb.2018.03.081
  [arXiv:1802.02716 [hep-ph]].
  %%CITATION = doi:10.1016/j.physletb.2018.03.081;%%
  %3 citations counted in INSPIRE as of 20 Dec 




%\cite{More:2017zqp}
\bibitem{More:2017zqp} 
  J.~More, A.~Mukherjee and S.~Nair,
  %``Wigner Distributions For Gluons,''
  Eur.\ Phys.\ J.\ C {\bf 78}, no. 5, 389 (2018)
  %doi:10.1140/epjc/s10052-018-5858-1
  [arXiv:1709.00943 [hep-ph]].
  %%CITATION = doi:10.1140/epjc/s10052-018-5858-1;%%
  %12 citations counted in INSPIRE as of 17 Jun 2019

%\cite{Kovchegov:2019rrz}
\bibitem{Kovchegov:2019rrz} 
  Y.~V.~Kovchegov,
  %``Orbital Angular Momentum at Small $x$,''
  JHEP {\bf 1903}, 174 (2019)
  %doi:10.1007/JHEP03(2019)174
  [arXiv:1901.07453 [hep-ph]].
  %%CITATION = doi:10.1007/JHEP03(2019)174;%%
  %1 citations counted in INSPIRE as of 06 Ma





%\cite{Efremov:1984ip}
\bibitem{Efremov:1984ip} 
  A.~V.~Efremov and O.~V.~Teryaev,
  %``QCD Asymmetry and Polarized Hadron Structure Functions,''
  Phys.\ Lett.\  {\bf 150B}, 383 (1985).
  %doi:10.1016/0370-2693(85)90999-2
  %%CITATION = doi:10.1016/0370-2693(85)90999-2;%%
  %335 citations counted in INSPIRE as of 06 Jun 2019

%\cite{Qiu:1991wg}
\bibitem{Qiu:1991wg} 
  J.~w.~Qiu and G.~F.~Sterman,
  %``Single transverse spin asymmetries in direct photon production,''
  Nucl.\ Phys.\ B {\bf 378}, 52 (1992).
  %doi:10.1016/0550-3213(92)90003-T
  %%CITATION = doi:10.1016/0550-3213(92)90003-T;%%
  %311 citations counted in INSPIRE a


%\cite{Braun:2009mi}
\bibitem{Braun:2009mi} 
  V.~M.~Braun, A.~N.~Manashov and B.~Pirnay,
  %``Scale dependence of twist-three contributions to single spin asymmetries,''
  Phys.\ Rev.\ D {\bf 80}, 114002 (2009)
  Erratum: [Phys.\ Rev.\ D {\bf 86}, 119902 (2012)]
  %doi:10.1103/PhysRevD.80.114002, 10.1103/PhysRevD.86.119902
  [arXiv:0909.3410 [hep-ph]].
  %%CITATION = doi:10.1103/PhysRevD.80.114002, 10.1103/PhysRevD.86.119902;%%
  %98 citations counted in INSPIRE as of 25 J



%\cite{Kang:2008ey}
\bibitem{Kang:2008ey} 
  Z.~B.~Kang and J.~W.~Qiu,
  %``Evolution of twist-3 multi-parton correlation functions relevant to single transverse-spin asymmetry,''
  Phys.\ Rev.\ D {\bf 79}, 016003 (2009)
  %doi:10.1103/PhysRevD.79.016003
  [arXiv:0811.3101 [hep-ph]].
  %%CITATION = doi:10.1103/PhysRevD.79.016003;%%
  %95 citations counted in INSPIRE


%\cite{Vogelsang:2009pj}
\bibitem{Vogelsang:2009pj} 
  W.~Vogelsang and F.~Yuan,
  %``Next-to-leading Order Calculation of the Single Transverse Spin Asymmetry in the Drell-Yan Process,''
  Phys.\ Rev.\ D {\bf 79}, 094010 (2009)
  %doi:10.1103/PhysRevD.79.094010
  [arXiv:0904.0410 [hep-ph]].
  %%CITATION = doi:10.1103/PhysRevD.79.094010;%%
  %82 citations counted in INSPIRE



%\cite{Pirnay:2013fra}
\bibitem{Pirnay:2013fra} 
  B.~M.~Pirnay,
  %``t3evol - Numerical Solution of Twist-three Evolution Equations,''
  arXiv:1307.1272 [hep-ph].
  %%CITATION = ARXIV:1307.1272;%%
  %1 citations counted in INSPIRE as of 29


%\cite{Ji:2012ba}
\bibitem{Ji:2012ba} 
  X.~Ji, X.~Xiong and F.~Yuan,
  %``Probing Parton Orbital Angular Momentum in Longitudinally Polarized Nucleon,''
  Phys.\ Rev.\ D {\bf 88}, no. 1, 014041 (2013)
  %doi:10.1103/PhysRevD.88.014041
  [arXiv:1207.5221 [hep-ph]].
  %%CITATION = doi:10.1103/PhysRevD.88.014041;%%
  %42 citations counted in INSPIRE as of 17 Jun 2019




    %\cite{Wakamatsu:2010qj}
\bibitem{Wakamatsu:2010qj} 
  M.~Wakamatsu,
  %``On Gauge-Invariant Decomposition of Nucleon Spin,''
  Phys.\ Rev.\ D {\bf 81}, 114010 (2010)
  %doi:10.1103/PhysRevD.81.114010
  [arXiv:1004.0268 [hep-ph]].
  %%CITATION = doi:10.1103/PhysRevD.81.114010;%%
  %98 citations counted in INSPIRE as of 06 Jun 2019

  
  %\cite{Ji:1996ek}
\bibitem{Ji:1996ek} 
  X.~D.~Ji,
  %``Gauge-Invariant Decomposition of Nucleon Spin,''
  Phys.\ Rev.\ Lett.\  {\bf 78}, 610 (1997)
  %doi:10.1103/PhysRevLett.78.610
  [hep-ph/9603249].
  %%CITATION = doi:10.1103/PhysRevLett.78.610;%%
  %1678 citations counted in INSPIRE 
  
  
  
  %\cite{Burkardt:2008ua}
\bibitem{Burkardt:2008ua} 
  M.~Burkardt and H.~BC,
  %``Angular Momentum Decomposition for an Electron,''
  Phys.\ Rev.\ D {\bf 79}, 071501 (2009)
  %doi:10.1103/PhysRevD.79.071501
  [arXiv:0812.1605 [hep-ph]].
  %%CITATION = doi:10.1103/PhysRevD.79.071501;%%
  %70 citations counted in INSPIRE as of 12 Jun 2019
  

  
  %\cite{Liu:2014fxa}
\bibitem{Liu:2014fxa} 
  T.~Liu and B.~Q.~Ma,
  %``Angular momentum decomposition from a QED example,''
  Phys.\ Rev.\ D {\bf 91}, 017501 (2015)
  %doi:10.1103/PhysRevD.91.017501
  [arXiv:1412.7775 [hep-ph]].
  %%CITATION = doi:10.1103/PhysRevD.91.017501;%%
  %9 citations counted in INSPIRE as of 12 Jun 2019
  
  
  %\cite{Ji:2015sio}
\bibitem{Ji:2015sio} 
  X.~Ji, A.~Schäfer, F.~Yuan, J.~H.~Zhang and Y.~Zhao,
  %``Spin decomposition of the electron in QED,''
  Phys.\ Rev.\ D {\bf 93}, no. 5, 054013 (2016)
  %doi:10.1103/PhysRevD.93.054013
  [arXiv:1511.08817 [hep-ph]].
  %%CITATION = doi:10.1103/PhysRevD.93.054013;%%
  %7 citations counted in INSPIRE as of 12 Ju
  
  %\cite{Engelhardt:2017miy}
\bibitem{Engelhardt:2017miy} 
  M.~Engelhardt,
  %``Quark orbital dynamics in the proton from Lattice QCD -- from Ji to Jaffe-Manohar orbital angular momentum,''
  Phys.\ Rev.\ D {\bf 95}, no. 9, 094505 (2017)
  %doi:10.1103/PhysRevD.95.094505
  [arXiv:1701.01536 [hep-lat]].
  %%CITATION = doi:10.1103/PhysRevD.95.094505;%%
  %20 citations counted in INSPIRE as of 12 Jun 2019
  
  
  %\cite{Engelhardt:2019lyy}
\bibitem{Engelhardt:2019lyy} 
  M.~Engelhardt, J.~Green, N.~Hasan, S.~Krieg, S.~Meinel, J.~Negele, A.~Pochinsky and S.~Syritsyn,
  %``Quark orbital angular momentum in the proton evaluated using a direct derivative method,''
  PoS SPIN {\bf 2018}, 047 (2019)
  %doi:10.22323/1.334.0115
  [arXiv:1901.00843 [hep-lat]].
  %%CITATION = doi:10.22323/1.334.0115;%%
  %2 citations counted in INSPIRE as of 12 Jun 2019
  
  
%\cite{Schafer:2012ra}
\bibitem{Schafer:2012ra} 
  A.~Schafer and J.~Zhou,
  %``A Note on the scale evolution of the ETQS function $T_F(x,x)$,''
  Phys.\ Rev.\ D {\bf 85}, 117501 (2012)
  %doi:10.1103/PhysRevD.85.117501
  [arXiv:1203.5293 [hep-ph]].
  %%CITATION = doi:10.1103/PhysRevD.85.117501;%%
  %30 citations counted in INSPI

%\cite{Ma:2012xn}
\bibitem{Ma:2012xn} 
  J.~P.~Ma and Q.~Wang,
  %``Scale Dependence of Twist-3 Quark-Gluon Operators for Single Spin Asymmetries,''
  Phys.\ Lett.\ B {\bf 715}, 157 (2012)
  %doi:10.1016/j.physletb.2012.07.036
  [arXiv:1205.0611 [hep-ph]].
  %%CITATION = doi:10.1016/j.physletb.2012.07.036;%%
  %36 citations counted in INSPIRE as of 06 May 


%\cite{Yoshida:2016tfh}
\bibitem{Yoshida:2016tfh} 
  S.~Yoshida,
  %``New pole contribution to $P_{h\perp}$-weighted single-transverse spin asymmetry in semi-inclusive deep inelastic scattering,''
  Phys.\ Rev.\ D {\bf 93}, no. 5, 054048 (2016)
  %doi:10.1103/PhysRevD.93.054048
  [arXiv:1601.07737 [hep-ph]].
  %%CITATION = doi:10.1103/PhysRevD.93.054048;%%
  %8 citations counted in INSPIRE as of 06 May


%\cite{Burkardt:2012sd}
\bibitem{Burkardt:2012sd} 
  M.~Burkardt,
  %``Parton Orbital Angular Momentum and Final State Interactions,''
  Phys.\ Rev.\ D {\bf 88}, no. 1, 014014 (2013)
  %doi:10.1103/PhysRevD.88.014014
  [arXiv:1205.2916 [hep-ph]].
  %%CITATION = doi:10.1103/PhysRevD.88.014014;%%
  %76 citations counted in INSPIRE as of 17 Jun 2019

  %\cite{Ji:1995cu}
\bibitem{Ji:1995cu} 
  X.~D.~Ji, J.~Tang and P.~Hoodbhoy,
  %``The spin structure of the nucleon in the asymptotic limit,''
  Phys.\ Rev.\ Lett.\  {\bf 76}, 740 (1996)
  %doi:10.1103/PhysRevLett.76.740
  [hep-ph/9510304].
  %%CITATION = doi:10.1103/PhysRevLett.76.740;%%
  %149 citations counted in INSPIRE as of 17 Jun 2019

  
  
  
\end{thebibliography}
\end{document}